\documentclass[twocolumn]{aastex63}

\def\fid{\texttt{fiducial}}
\def\ldcloud{\texttt{ldcloud}}
\def\ldwind{\texttt{ldwind}}
\def\swind{\texttt{swind}}
\def\fwind{\texttt{fwind}}

\def\NW{\texttt{NW}}

\received{date}
\revised{date}
\shorttitle{Cloud Crushing with Star Formation}
\shortauthors{Tonnesen et al.}
\graphicspath{{./}{figures/}}

\begin{document}

\title{Great Balls of Fire: Star Formation in Gas Clouds Accelerated by a Hot Wind}

\author[0000-0002-8710-9206]{Stephanie Tonnesen}
\affiliation{Center for Computational Astrophysics, Flatiron Institute, 
162 5th Avenue,
New York, NY 10010, USA}
\author[0000-0001-5303-6830]{Rory Smith}
\affiliation{Departamento de Física, Universidad Técnica Federico Santa María, 
Vicuña Mackenna 3939, 
San Joaquín, Santiago de Chile, Chile}
\affiliation{Millennium Nucleus for Galaxies (MINGAL),
Valparaiso, Chile}
\author[0000-0001-8751-8360]{Bianca Poggianti}
\affiliation{INAF - Osservatorio Astronomico di Padova,
Vicolo dell’Osservatorio 5, I-35122 Padova, Italy
}
\author[0000-0003-0980-1499]{Benedetta Vulcani}
\affiliation{INAF - Osservatorio Astronomico di Padova,
Vicolo dell’Osservatorio 5, I-35122 Padova, Italy
}
\author[0000-0001-7011-9291]{Nina Akerman}
\affiliation{INAF - Osservatorio Astronomico di Padova,
Vicolo dell’Osservatorio 5, I-35122 Padova, Italy
}
\affiliation{Dipartimento di Fisica e Astronomia ‘Galileo Galilei’, Universit `a di Padova, vicolo dell’Osservatorio 3, I-35122 Padova, Italy}
\author[0000-0002-7296-9780]{Marco Gullieuszik}
\affiliation{INAF - Osservatorio Astronomico di Padova,
Vicolo dell’Osservatorio 5, I-35122 Padova, Italy
}
\author[0000-0002-3818-1746]{Eric Giunchi}
\affiliation{Dipartimento di Fisica e Astronomia "Augusto Righi", Università di Bologna, via Piero Gobetti 93/2, I-40129 Bologna, Italy}
\author[0000-0002-5655-6054]{Antonino Marasco}
\affiliation{INAF - Osservatorio Astronomico di Padova,
Vicolo dell’Osservatorio 5, I-35122 Padova, Italy
}
\author[0000-0003-1581-0092]{Alessandro Ignesti}
\affiliation{INAF - Osservatorio Astronomico di Padova,
Vicolo dell’Osservatorio 5, I-35122 Padova, Italy
}
\correspondingauthor{Stephanie Tonnesen}
\email{stonnesen@flatironinstitute.org}

\begin{abstract}

Satellite galaxies undergo ram pressure stripping, in which their gas is directly removed by a hydrodynamical interaction with the surrounding host halo gas. In clusters, ram pressure stripped tails of gas have been observed to be multiphase, even forming stars within the stripped material.  Some observations find a specific age gradient along the tail, with old stars closer to the galaxy disk, and a ``fireball'' toy model has been proposed in which a gas cloud being accelerated away from a galaxy continuously forms stars.  In this paper, we simulate individual gas clouds (with masses $\sim$10$^6$ M$_{\odot}$ and radii of a few-100 pc) interacting with an intracluster medium wind, and include star formation.  We find that our accelerating clouds do generally produce a stellar age gradient with younger stars formed farther along the wind direction and with higher velocities.  However, our simulations are more physically accurate than an empirical model of monolithic cloud acceleration, leading to strongly nonmonotonic age gradients.  First, the evolution of the gas cloud, both from cloud compression and collapse as well as from the shredding of cloud material into downwind filaments, can lead to stars formed simultaneously at a range of heights and velocities.  Second, the gravity from the gas and stars of the cloud can lead to velocity evolution of newly-formed stars.  We conclude that the most distinct fireball stellar age gradients are formed from star-forming clouds that are rapidly accelerated and shredded by their surroundings.   

\end{abstract}



\section{Introduction} \label{sec:intro}

When satellite galaxies orbit within a massive cluster, their interstellar medium (ISM) can be directly removed via a hydrodynamical interaction with the intracluster medium (ICM) called ram pressure stripping \citep{gunn_gott_1972}.  This is generally considered to be a ubiquitous mechanism that acts on many, if not most, galaxies in clusters \citep{boselli_environmental_2006, Boselli_2016A&A...596A..11B,Vulcani_ubiquitousRPS_2022ApJ...927...91V}.  There now exists a large observational body of work showing long unilateral gas tails that are considered the main signature of ram pressure stripping.% 

These stripped tails of gas have been observed at a large range of wavelengths.  X-ray emission has been observed in the tails of stripped massive ellipticals \citep{Kim_xray_2008ApJ...688..931K,Machacek_Xrayelliptical_2006ApJ...644..155M,Wang_Xray_2004ApJ...611..821W} as well as in the likely mixed ISM-ICM in tails behind sub-L$^*$ galaxies \citep{Sun_2022_ICMISMtails}.   Optical emission has also been observed from gas within tails \citep{poggianti_gaspI_2017,Yagi_opticalfilament_2007ApJ...660.1209Y, Fumagalli_MUSE_ESO137_2014MNRAS.445.4335F,Boselli_VESTIGEIV_2018A&A...620A.164B,Poggianti_2025A&A_GASPsummarypaper}. Moving to longer wavelengths, tails have been observed in radio continuum observations from 100 MHz up to a few GHz \citep{Gavazzi_1987A&A...186L...1G,Gavazzi_1995A&A...304..325G, Ignesti_2023A&A_RPSradiocontinuum,Chen_2020MNRAS_RPSradiotails, Ignesti_2022ApJ_GASP_LOFAR,roberts_lotss_2021-1,roberts_lotss_2022,roberts_lotss_jellyfishradiotails_2021}, as well as in HI emission \citep{Kenney_HI4522_2004AJ....127.3361K,deb_gasp_2022,ramatsoku_gaspHI_2020,oosterloo_NGC4388_2005,Chung_VIVAsurvey_2007ApJ...659L.115C}.  More recently, ram pressure stripped tails have been observed at millimeter wavelengths \citep{Jachym_2017_D100,cramer_D100_2019,Sivinandam_molecularHESO137_2010ApJ...717..147S, Moretti_2018_GASPAPEX,Moretti_2020_JW100,moretti_GASPmolecular_2020}.  Indeed, multiwavelength observations exist for several individual tails, highlighting that multiphase gas exists in stripped tails in the ICM out to large distances \citep[eg][]{Poggianti_2019_GASPbaryoncycle,Sun_alsoSFESO137_2010ApJ...708..946S,Jachym_2019_ESO137,Kenney_2015_NGC4921,Owen_UVtail_2006AJ....131.1974O}.  The survival of multiphase gas within stripped tails has also been found in simulations, particularly high-resolution wind-tunnel simulations that can resolve the tail gas and allow radiative cooling to sub-10$^4$ K temperatures \citep[e.g.][]{kapferer_RPS_2009,tonnesen_tail_2010,Tonnesen_2011_Xraytails,lee_RPS_2020,lee_RPSISM_2022}.

Not only has gas been observed in ram pressure stripped tails, but stars and star-forming regions have also been found.  Early observations found stars quite close to the stripped galaxies \citep{Cortese_HIIregionRPS_2004A&A...416..119C, Kenney_Koopmann_HIINGC4522_1999AJ....117..181K}.  However, a large body of literature now shows that stars can form large distances from the stripped galaxies, with some long trails identified in targeted observations of known stripped galaxies \citep{sun_haESO137_2007,Yoshida_2008_fireballobs,Yagi_HIIregions_Virgotails_2013ApJ...778...91Y,Cortese_opticaltails_2007MNRAS.376..157C,Kenney_2014_fireballs}, and others found in larger surveys including galaxies in one or several clusters \citep{Gullieuszik_SFclumpsGASP_2023ApJ...945...54G, poggianti_gaspSFoutsidegalaxies_2019,owers_shockingtails_opticalobs_2012,ebeling_jellyfish_2014,Boselli_VESTIGE_SFtail_2018A&A...615A.114B}.  The previous works used H$\alpha$ emission to map star formation (SF), but UV observations have also identified SF in stripped material \citep{ Rawle_UV_IR_tailSF_2014MNRAS.442..196R, George_JO201UV_2018MNRAS.479.4126G, George_GASPUV_2023MNRAS.519.2426G,George_2025A&A_RPSUV}.  Simulations have also produced SF in ram pressure stripped tails, both in idealized wind-tunnel simulations of single galaxies \citep{Tonnesen_2012_RPSSFR,roediger_SFRPS_2014,lee_RPSISM_2022,kapferer_RPS_2009, Akerman_SFtails_2025A&A...698A.151A} (but see \citet{steinhauser_RPSsimulations_2016} for simulations without SF in the tails) and in cosmological simulations \citep{Goller_TNG_SFRPS_2023MNRAS.525.3551G}.    

\begin{figure}
    \centering
    \includegraphics[scale=0.56]{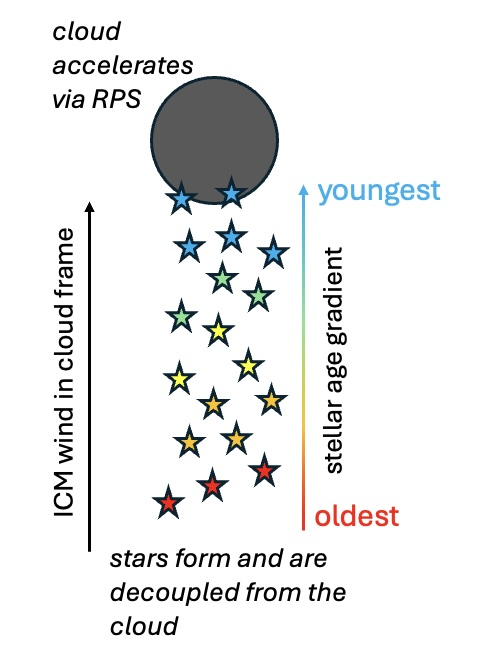}
    \caption{An illustration of a classic ``fireball''.  A gas cloud (gray sphere) is monolithically accelerated via an interaction with the surrounding gas halo, and forms stars along its path.  Stars that form are no longer accelerated and move ballistically.  This results in an age gradient with older stars upwind of younger stars, and the gas cloud downwind of the entire stellar population.  Although it is not modeled in this work, the stripped galaxy would be upwind of this cartoon.}
    \label{fig:fireball_illustration}
\end{figure}

Because stars should not be removed directly from a galaxy's disk through a hydrodynamical interaction with the ICM, these stars must have formed within the stripped material.  In order to explain observations of star-forming regions in stripped tails, a toy ``fireball'' star-formation scenario has been developed, which we illustrate in Figure \ref{fig:fireball_illustration} and can also be found in \citet{Jachym_2019_ESO137,Kenney_2014_fireballs}.  A fireball, or clump of tail stars whose ages decrease with increasing distance from the stripped galaxy, is formed from a dense gas cloud that is being accelerated away from a ram pressure stripped galaxy (in the galaxy frame) while forming stars. Although the dense cloud will continue to be accelerated until it reaches the ICM wind velocity, as soon as a star is formed it decouples from the cloud and the ICM wind.  This results in the cloud ``leaving behind'' a trail of stars, and those stars will show an age gradient that is younger closer to the accelerating cloud and older closer to the galaxy disk.

This straightforward toy model generally reproduces several observations comparing the spatial distribution of H$\alpha$ and UV emission as proxies for young and older stellar populations \citep[e.g.][]{Kenney_2014_fireballs, Smith_2010_Comafireballs,Giunchi_2023ApJ...949...72G,Giunchi_2023_cloudmorphology,Werle_SFtails_2024A&A...682A.162W}, and using molecular gas observations to determine the relative position of the dense star-forming gas \citep[e.g.][]{Jachym_2017_D100,Jachym_2019_ESO137,Poggianti_2019_GASPbaryoncycle}.  Several high-resolution observations of spatially-correlated star-forming regions have found UV emission detected closer to the galaxy than H$\alpha$ emission and gas detected the farthest from the galaxy, in agreement with the fireball model. However, there are also observations that indicate that this is too simplified to explain all SF in stripped tails.  For example, using H$\alpha$ and UV observations of star-forming clumps in optical stellar complexes, \citep{Giunchi_2023_cloudmorphology} find that only a small majority ($\sim$60\%) of young star-forming complexes show the expected fireball age gradient.  Expanding the fireball scenario, \citet{Jachym_2019_ESO137} included the ablation of the cold star-forming cloud in their ``double-sided'' fireball model because in addition to H$\alpha$ emission observed between the dense gas cloud and the disk from newly formed stars, diffuse H$\alpha$ emission is observed to originate near the dense gas but extend away from the galaxy.  There are currently very few simulations of individual star-forming clouds in the ICM,  but \citet{Calura_SECCO1_2020MNRAS.499.5873C} simulate a massive (10$^7$ M$_\odot$) low-density cloud that is a few kpc across to model SECCO 1, a star-forming cloud observed far from any galaxy in the Virgo cluster.  When they include SF, the age distribution generally follows the fireball prediction, with the oldest stars upwind of the younger stars and the gas the farthest downwind. 

Simulations without SF in small cold clouds evolving in a hot surrounding wind have recently proliferated, generally in the context of cold clouds in the CGM of Milky Way-like galaxies \citep[e.g.][]{GronkeOh2018_coldcloudgrowth,GronkeOh2020a_coldgasentrains,Armillotta2016_diskcorona,Marinacci2010_coldgasaccretion,Abruzzo_2022ApJ...925..199A}.  Importantly, they find that radiative cooling within the mixing layer between the cold and hot gas is necessary for cloud survival. Although the details of the criteria vary, the critical overarching theme is that when the cooling rate is faster than the cloud destruction rate not only does a cloud survive, but it can grow in mass \citep{GronkeOh2018_coldcloudgrowth,Li_Hopkins_2020MNRAS_cloudsurvival_with_conductionviscosity,Sparre_2020MNRAS_cloudcrushingMHD,Abruzzo_2022ApJ...925..199A}.  In contrast to a simple fireball picture in which a monolithic cloud is accelerated by a wind, these simulations have shown that originally spherical uniform clouds become shredded into long filamentary structures that are able to cool and accrete cold gas all along their length.  

When one considers the simulation findings that clouds can be constantly destroyed and reformed as stretched filaments populated by cold clumps, it becomes less clear that the straightforward fireball picture of a single cloud accelerating as it forms stars should describe the evolution of cold clouds of gas in a ram pressure stripped tail.  In this work we run a suite of simulations of small (r $\sim$200 pc) gas clouds within a hot and fast ICM wind including radiative cooling and SF to focus directly on the stellar populations that can form from stripped clouds that are surrounded by an ICM wind.  

In Section \ref{sec:method} we introduce our simulation code and the suite of runs analyzed in this paper.  We then present the differences in the global star formation rate across the simulations in Section \ref{sec:SFR}.  In Section \ref{sec:results_stellardist} we begin to look at the stellar distribution in detail in our simulations, beginning with visual inspections (Sections \ref{sec:results_fidstellardist} \& \ref{subsec:results_viscompstellardist}) and then quantitatively comparing the stellar age distributions (Section \ref{subsec:results_quantcompstellardist}).  We then dive into what drives the stellar distribution in Section \ref{sec:gasandstars}, starting with the initial velocity and height distribution (Section \ref{sec:zhists}), and continuing with how the stellar velocity changes over time (Section \ref{sec:stellarvelocityevolution}). In Section \ref{sec:discussion} we discuss the circumstances under which our simulations form fireball-like stellar clumps and how our simulations compare to the observed populations of stars formed in stripped tails.  Finally, we conclude (Section \ref{sec:conclusion}) by summarizing our results.  

\section{Method} \label{sec:method}

We run a suite of three-dimensional cloud `wind-tunnel' simulations\footnote{`wind-tunnel' simulations are named for the gas that flows through the entire simulated region, in our case inflowing through the -z side and outflowing through the +z side of the box.} using the adaptive mesh refinement (AMR) code Enzo \citep{Bryan_2014_Enzo}. The cloud is placed in a $25^{3}$ kpc$^{3}$ simulation volume with a $256^{3}$ root grid resolution. We allow up to six levels of refinement so that the highest spatial resolution is $1.5$ pc. Given that our cloud radius is at least 215 pc, this leads to more than 140 cells across the cloud diameter. We note that our choice of a cubic domain, in which very little of the total volume ever interacts with the cloud material, allows us to disregard any boundary effects (such as spurious reflections) while not slowing the computation because our cost is completely dominated by the refined cloud material. We use the Grackle chemistry and cooling library\footnote{\url{https://grackle.readthedocs.io/}} to model radiative cooling \citep{smith_grackle_2017}, which calculates photoheating and photoionization from the UV background of \cite{haardt_UVB_2012}. We have applied the Grackle self shielding method ``3'', in which analytic fits to radiative transfer simulations published in \citet{Rahmati_2013_selfshielding}are applied independently on each computational element.  This method attenuates the ionization and photoheating rates in both neutral H and neutral He, while ignoring He$^+$ photoionization and photoheating from the UV background entirely.  Future careful tests of this choice would allow us to determine the accuracy of this approximation \citep[although we note that this is recommended as the most reasonable choice in ][]{smith_grackle_2017}.

We use the SF recipe of \cite{goldbaum_selfgravdisks_2015} and the stellar and supernovae feedback model from \cite{goldbaum_SFfeedbackdisk_2016}.  In all simulations star particles can be formed once a gas cell at the highest refinement level has reached the Jeans criterion with a density threshold of 100 cm$^{-3}$, and SF proceeds with an efficiency of 2\%. This density threshold is in good agreement with similarly-resolved simulations \citep[e.g.][]{Hopkins_FIRE1_2014MNRAS.445..581H, Kannan_THESANzoom_2025arXiv250220437K, Marinacci_SMUGGLE_2019MNRAS.489.4233M}, and the SF efficiency in these simulations ranges from 1\% \citep{Marinacci_SMUGGLE_2019MNRAS.489.4233M} to 100\% \citep{Hopkins_FIRE1_2014MNRAS.445..581H, Kannan_THESANzoom_2025arXiv250220437K}.  
In practice in our simulations, the cell mass is not larger than the Jeans mass until the density is more than an order of magnitude larger than our threshold, and when we use a threshold of 1500 cm$^{-3}$, the SF history is qualitatively unchanged (see Section \ref{sec:disc_caveats} for discussion).  The minimum star particle mass is 10 M$_{\odot}$.  

The formed star particles then deposit energy into the gas in the form of
stellar and supernovae feedback, which includes the terminal momentum input from the number of supernovae
expected to go off during a given timestep, adding any
additional energy in the form of thermal energy \citep{goldbaum_SFfeedbackdisk_2016}.

We use yt, a multi-code toolkit for analyzing and visualizing astrophysical simulation data \citep{Turk_yt_2011ApJS}, for most of the analysis in this work.

\subsection{The Simulation Suite}

In this paper we present six wind-tunnel simulations in which we follow a dense gas cloud embedded in a hotter surrounding medium for 100 Myr.  In each simulation we vary either the initial cloud density profile or the wind properties (density or velocity of the surrounding hot gas flows). The parameters describing the simulations are listed in Table \ref{tab:clouds}.  All simulations are named based on how they differ from our {\fid} simulation of a cloud that is hit by an oncoming ICM wind.  In detail, {\fwind} stands for the fast wind, {\swind} for the slow wind, and {\ldwind} for the low density wind.  We also have a run that is initialized in a static medium with the same pressure as the wind in {\fid}: {\NW}, which stands for no wind and as such evolves in a static medium.  Finally, {\ldcloud} has the same wind as in {\fid}, but the cloud initially has a lower density and a larger radius such that the total gas cloud mass is the same as in the other simulations.  

\subsubsection{Cloud Initial Conditions}\label{sec:cloud_initial}
%\vspace{-2.5em}
\begin{deluxetable*}{cccccccc}\label{tab:clouds}
\tablecaption{Cloud Simulations} 
%\tablewidth{0pt}
\tablehead{\colhead{identifier} & \colhead{$M_{\rm gas,init}$} & \colhead{$r_{\rm gas,init}$} & \colhead{$\rho_{\rm wind}$} & \colhead{$v_{\rm wind}$} & \colhead{$T_{\rm wind}$} &  \colhead{$P_{\rm therm, wind}$}  & \colhead{$M_{\rm fin,stars}$} \\
\colhead{ } & \colhead{(10$^5$ M$_{\odot}$)} & \colhead{(pc)} & \colhead{(10$^{-27}$ g cm$^{-3}$)} & \colhead{(10$^3$ km s$^{-1}$)} & \colhead{(10$^7$ K)} &  \colhead{(10$^{-11}$dyne cm$^{-2}$)} & \colhead{(10$^5$ M$_{\odot}$)} \\
\colhead{(1)} & \colhead{(2)} & \colhead{(3)} & \colhead{(4)} & \colhead{(5)} & \colhead{(6)} & \colhead{(7)} & \colhead{(8)}}
\startdata
\fid	&	7.8	&	215	&	10	&	1	&	3.4	&	4.7	&	1.1 \\
\ldcloud	    &	7.8	    &	464 	&	10	&	1    &	3.4	&	4.7	&	0.2 \\
\ldwind	    &	7.8	    &	215	 &	5.6	&	1	&	3.4	&	2.6	& 1.5 \\
\swind	    &	7.8    &	215	&	10	&	0.77	&	2.3	&	3.0	& 1.1 \\
\fwind	    &	7.8	    &	215	&	10	&	1.5	&	6.3	&	8.85 &	0.9 \\
\NW	    &	7.8	    &	215	&	10	&	0.0 	&	3.9	&	5.5	& 7.3 \\
\enddata
%\tablecomments{}
\end{deluxetable*}

In all simulations, the gas cloud is initially at [0.5,0.5,0.15] in code units ([12.5,12.5,3.75] kpc in the [25,25,25] kpc box).  The placement near the bottom of our simulated box (low $z$ position) allows us to follow the accelerated cloud and stripped material for more than 20 kpc (several tens of the initial cloud radius).  The simulations are based on the galaxy-scale wind tunnel simulations in \citet{Tonnesen_2009_RPSdisks}, although in these simulations the clouds are purely gas (with small non-zero mass in a stellar and dark matter static potential consisting of less than 0.001\% of the total cloud mass for code stability).  
 
All of the clouds have the same initial gas mass of $\sim$8$\times$10$^5$ M$_{\odot}$ (as listed in Table \ref{tab:clouds}).  We produce a spherical cloud by setting both the vertical and cylindrical radial gas density scale lengths to 86 pc (186 pc for {\ldcloud}).  The cloud is manually truncated at 215 pc (464 pc for {\ldcloud}).  The cloud temperature profile is then calculated so that the initial condition of the cloud is in pressure equilibrium with its surroundings.  This mass and radius selection for our clouds results in stellar masses and radii of the main stellar clump that are near the peak of the observed distribution of stellar clump masses and radii in \citet{Giunchi_2023_cloudmorphology, Werle_SFtails_2024A&A...682A.162W, Giunchi_2025_GASP_clump_mass_function}, as we will mention in later sections. 

\begin{figure}
    \centering
    \includegraphics[scale=0.55,trim={0mm 2mm 0mm 10mm},clip]{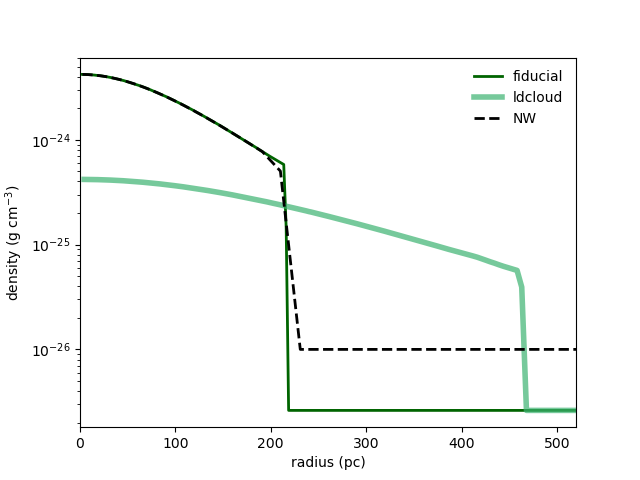}\\
    \includegraphics[scale=0.55,trim={0mm 2mm 0mm 10mm},clip]{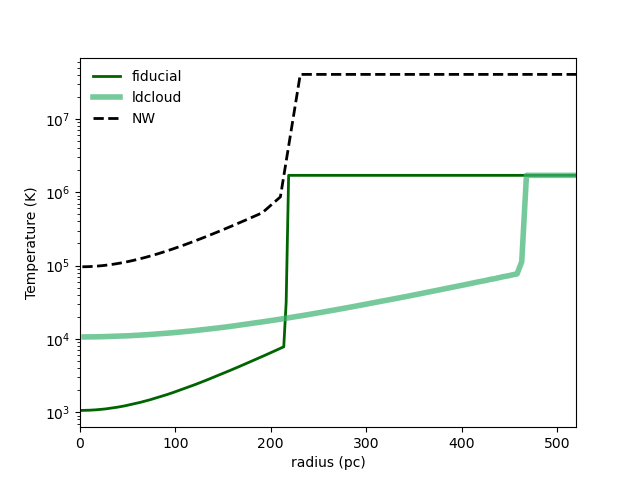}
    \caption{The initial cloud temperature and density for the {\NW}, {\fid} (which is the same for all the wind runs except {\ldcloud}), and {\ldcloud} runs. }
    \label{fig:HM_cloudinits}
\end{figure}

In Figure \ref{fig:HM_cloudinits} we show the initial density and temperature profiles for the clouds \footnote{We note that as the simulations include radiative cooling and allow cold and hot gas to mix, there forms a cocoon of $\sim$10$^4$ K gas around the cold cloud (and later around the cold fragmented tail material), in agreement with other cloud crushing simulations \citealp[e.g.][]{GronkeOh2020a_coldgasentrains}.}.  We do not show a thermal pressure profile as the clouds are initialized to be in pressure equilibrium with the surroundings.  Most clouds follow the {\fid} cloud profile, with only the {\ldcloud} and {\NW} having different initial conditions.  The {\ldcloud} has a lower density because it is larger but has the same mass as the {\fid} cloud, and therefore it is also at a higher temperature so that it is in pressure equilibrium with the low density initial surroundings.  The {\NW} cloud has nearly the same density profile as the {\fid} cloud, but has a higher temperature because it is initially surrounded by higher pressure gas.  This higher pressure gas is selected to match the thermal pressure of the hot high-density wind that impacts all the other clouds. This allows us to compare the SF in a cloud surrounded by a hot \textit{wind} to the SF in a cloud surrounded by hot \textit{static} gas with the same thermal pressure.

\subsubsection{Wind Parameters}\label{sec:method_runs}

Although there is little variation in the cloud properties across most of our simulation suite (see Figure \ref{fig:HM_cloudinits} and the discussion in Section \ref{sec:cloud_initial}), we vary the wind properties. In all cases except for {\NW}, the inflow enters from the $-z$ box face and flows in the $+z$ direction.  The inflow begins immediately, and reaches the clouds between 1.7 Myr ({\fwind}) and 3.1 Myr ({\swind}) from the beginning of the simulation (the {\NW} cloud is initialized while surrounded by the high pressure gas listed in the table).  Each inflow is constant with the parameters listed in Table \ref{tab:clouds}.  

So that we have a clean jump between the pre-wind and wind surroundings, we choose the inital pre-wind ICM density and temperature such that the wind will move through the box with a Mach number of 2.  This method directly follows that in several papers studying ram pressure stripping \citep[e.g.][]{Tonnesen_2009_RPSdisks, Tonnesen_2010_multiphasetails, Tonnesen_2011_Xraytails, Tonnesen_2012_RPSSFR, akerman_2023_gasflowsinRPS, Zhu_Tonnesen_2024_SFinRPSdwarfs}.  We choose to have the wind flow through a boundary rather than simply assigning the surrounding gas an initial velocity.  One reason for this choice is that if all the hot gas was initially moving, there would be a low pressure region downwind of the cloud that would unrealistically deform and accelerate cloud material.  From a physical scenario point of view, we also expect that recently-stripped clouds will be rapidly moving into higher density regions (as they continue to plunge through the ICM) that will result in increasing ram pressure and thermal pressure around the cloud.  Although this would not necessarily result in a sudden shock crossing the cloud, because there is turbulence in stripped tails this is not an unreasonable scenario \citep[e.g.][]{Fumagalli_MUSE_ESO137_2014MNRAS.445.4335F}.  Future work varying the initial inflow could test the impact of this choice.

\section{Global Star Formation Rate}\label{sec:SFR}

\begin{figure}
    \centering
    \includegraphics[scale=0.56]{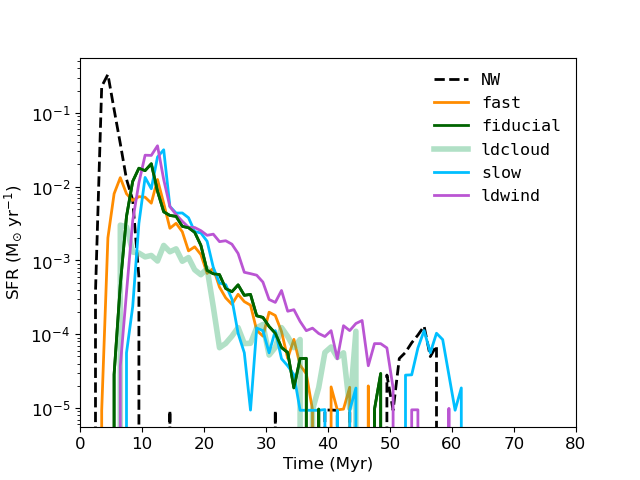}
    \caption{The global star formation rate (SFR) over time in our main suite of simulations. }
    \label{fig:HM_SFRs}
\end{figure}

We first consider whether the wind or cloud properties affect the total stellar mass and global star formation rate (SFR) in the simulations.

In Table \ref{tab:clouds} we report the final total stellar mass and in Figure \ref{fig:HM_SFRs} we plot the global SFR within the entire simulated region in the bottom panel. We note that this range of stellar masses is within the range of masses in the observed stellar clumps outside ram pressure stripped galaxies in \citet{Giunchi_2023_cloudmorphology} and \citet{Werle_SFtails_2024A&A...682A.162W}, and indeed in those works most of the stellar clumps have masses near 10$^5$ M$_{\odot}$.  However, we also note that SF in our simulations lasts for less than the 100 Myr of our full simulation run-time, which is shorter than the age distributions in many of the stellar clumps identified in those works. Although we will compare our simulations to observations in Section \ref{sec:disc_observations}, here we focus on comparing between simulations.

The most striking difference between the simulations is that the {\NW} simulation forms the most stars (Table \ref{tab:clouds}) and reaches the highest SFR (Figure \ref{fig:HM_SFRs}).  Recall from Figure \ref{fig:HM_cloudinits} that the NW galaxy is initially surrounded by a static but high thermal pressure medium, which sets the cloud temperature to $\sim$10$^5$ K.  This gas rapidly cools, with a cooling time of about 0.003 Myr, and is then compressed such that the density increases to above the SF threshold of 100 cm$^{-3}$ within about 3 Myr.  At this point the cloud quickly turns nearly all of its gas into stars.  On the other hand, while the clouds in the wind runs also rapidly cool, with cooling times of less than 0.05 Myr, because they are surrounded by lower pressure gas they are compressed more slowly, allowing for an interaction with the inflowing high-pressure ICM before star formation occurs (Figure \ref{fig:HM_cloudinits}). Once the wind hits the {\fid} cloud the surrounding pressure rapidly jumps by a factor of about 60 to match that of {\NW} ({\ldwind} and {\fwind} by factors of 30 and 105, respectively), which will more rapidly compress the clouds to densities above the SF limit.  However, the wind cases form about a factor of ten fewer stars than {\NW}, which is likely because in the wind cases the surrounding high thermal pressure gas also has a high relative velocity to the cloud.  This indicates that the SF from the radiative cooling and collapse of our simulated cloud is not enhanced by a compressive shock front from a hot wind, and instead is reduced due to the interaction with the ICM wind. %This gas rapidly cools and the density increases to above the SF threshold of 100 cm$^{-3}$.  At this point the cloud quickly turns nearly all of its gas into stars.  On the other hand, the clouds in the wind runs do not rapidly cool and collapse before they interact with the inflowing high-pressure ICM, as they are already at low temperatures with longer cooling times to be in pressure equilibrium with the initial lower-density surrounding medium (Figure \ref{fig:HM_cloudinits}). Once the wind hits the {\fid} cloud the surrounding pressure rapidly jumps by a factor of about 60 to match that of {\NW} ({\ldwind} and {\fwind} by factors of 30 and 105, respectively), pushing the cloud gas density in all but {\ldcloud} above the SF limit.  However, the wind cases form about a factor of ten fewer stars than {\NW}, which is likely because in the wind cases the surrounding high thermal pressure gas also has a high relative velocity to the cloud.  This indicates that the SF from the radiative cooling and collapse of our simulated cloud is not enhanced by a compressive shock front from a hot wind, and instead is reduced due to the interaction with the ICM wind.  

Another difference highlighted in Table \ref{tab:clouds} and Figure \ref{fig:HM_SFRs} is that the {\ldcloud} gas cloud, which contains the same amount of mass as the {\fid} cloud but has a radius that is $\sim$2.15 times larger, forms fewer stars and reaches a lower peak SFR than any runs with the higher density cloud.  Notably, at the {\ldcloud} density, the cooling time becomes about 1.5 Myr at 10$^3$ K, so the cloud does not rapidly become as strongly underpressured as the other clouds.  Further, not only will a density enhancement by a factor of 60 not push the {\ldcloud} initial density (Figure \ref{fig:HM_cloudinits}) above the SF limit, but compression-induced central collapse will take longer due to the larger cloud size.  Therefore, both compression and radiative cooling-driven collapse will take longer to push the {\ldcloud} over the SF density threshold.  This results in less rapid SF and more lower-density gas being accelerated and stretched by the wind once the high pressure ICM wind has surrounded the {\ldcloud} cloud in comparison to the {\fid} cloud.  Although the initial SF peak of {\ldcloud} is lower than the other wind runs, at later times, beyond about 20 Myr, the SFR is quite similar between the {\ldcloud} and {\fid} simulations (which are impacted by the same wind, see Table \ref{tab:clouds}).

The stripped clouds show quite similar SFR and stellar mass evolution with an early SFR peak followed by a steeply decaying SFR that is reflected in the majority of stellar mass forming in the simulation within the first $\sim$10 Myr.  However, we also note that there are minor differences between the total stellar mass and global SFRs in the runs in which we vary the (constant) wind parameters.  As discussed in Section \ref{sec:method_runs}, the strength of ram pressure increases from the {\swind} run (and the equal ram pressure {\ldwind} run) to the {\fid} run, and is highest in the {\fwind} run.  We find that increased ram pressure results in fewer stars formed and a slightly lower SFR peak.  This agrees with the result that {\NW} has the largest stellar mass and highest peak SFR.  Of the wind runs, the {\ldwind} run forms the most stars.  Interestingly, this is not due to an enhancement in the peak SFR reached by {\ldwind}, but rather to the somewhat enhanced SFR extending for about forty Myr after the peak. 

In summary, in our simulations including a wind interacting with a cloud reduces the total cloud SF and the peak SFR by about 90\%.  When comparing the wind runs, although the differences are small, there is some indication that reducing the ram pressure of the wind allows for more SF (comparing {\ldwind}, {\fid}, and {\fwind}).  Finally, a wind impacting a larger, lower density cloud of the same total mass ({\ldcloud}) leads to the least SF.

\section{The Stellar Age Distribution in Stripped Clouds}\label{sec:results_stellardist}

In this section we look in detail at the age distribution in our cloud simulations.  In Figures \ref{fig:projections} - \ref{fig:1p6projs_comps} we show a series of snapshots from our simulations to gain visual intuition on the evolution of the stellar morphology in different runs.  We will then quantitatively examine the age distribution of stars formed from the cloud material as a function of height (in the $z$-direction, which is the wind direction) across our simulations in Section \ref{subsec:results_quantcompstellardist}.

\subsection{A Visual Inspection of the Fiducial Simulation}\label{sec:results_fidstellardist}

\begin{figure*}
    \centering
    \includegraphics[scale=0.19,trim={0mm 0mm 52mm 0mm},clip]{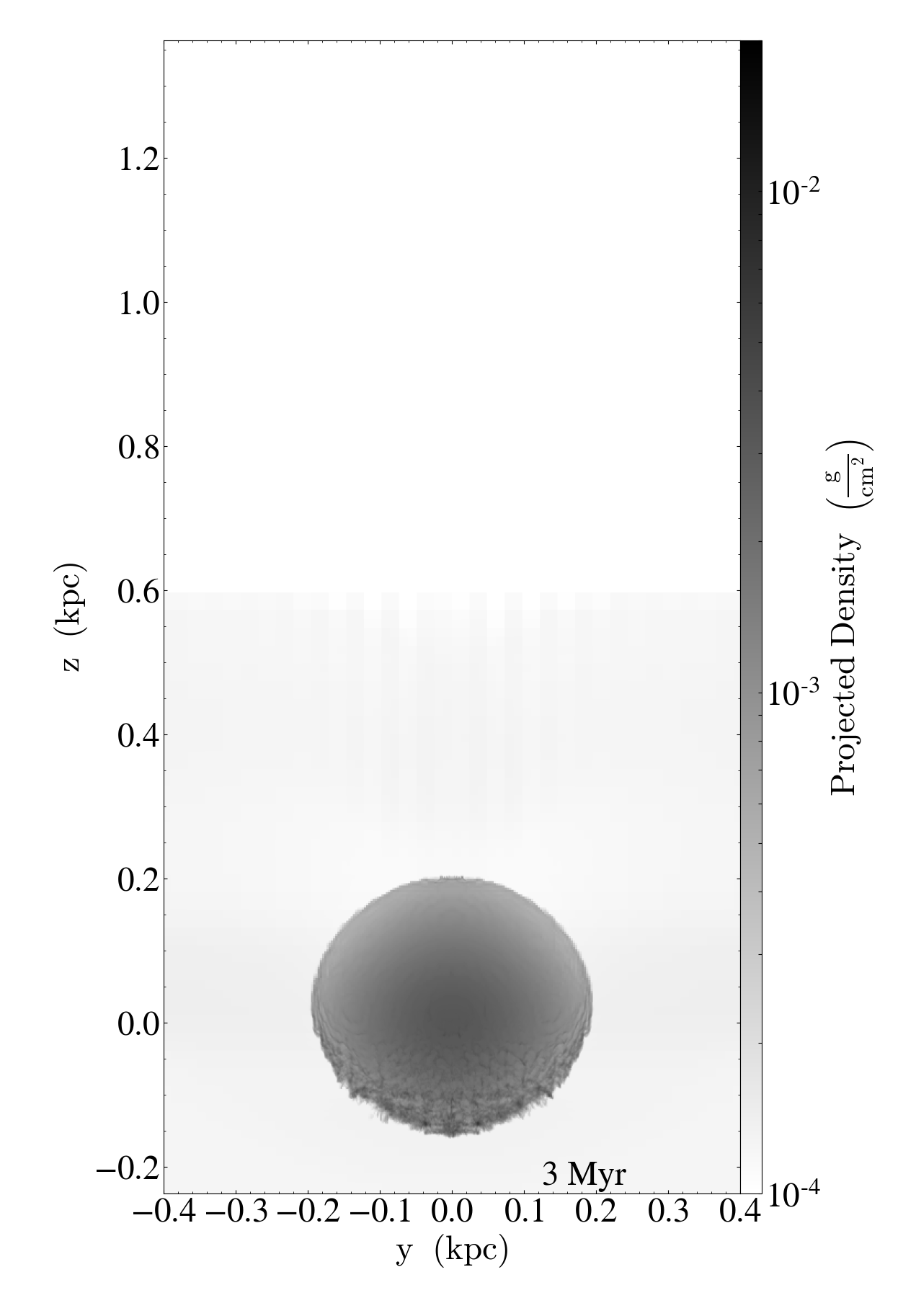}
    \includegraphics[scale=0.38,trim={0mm 0mm 36mm 0mm},clip]{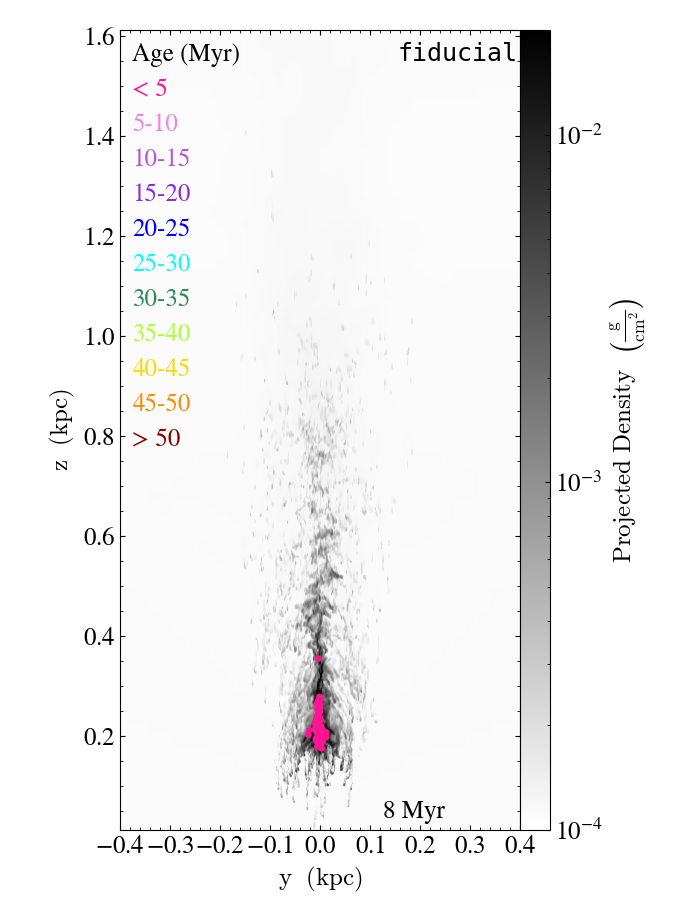}
    \includegraphics[scale=0.38]{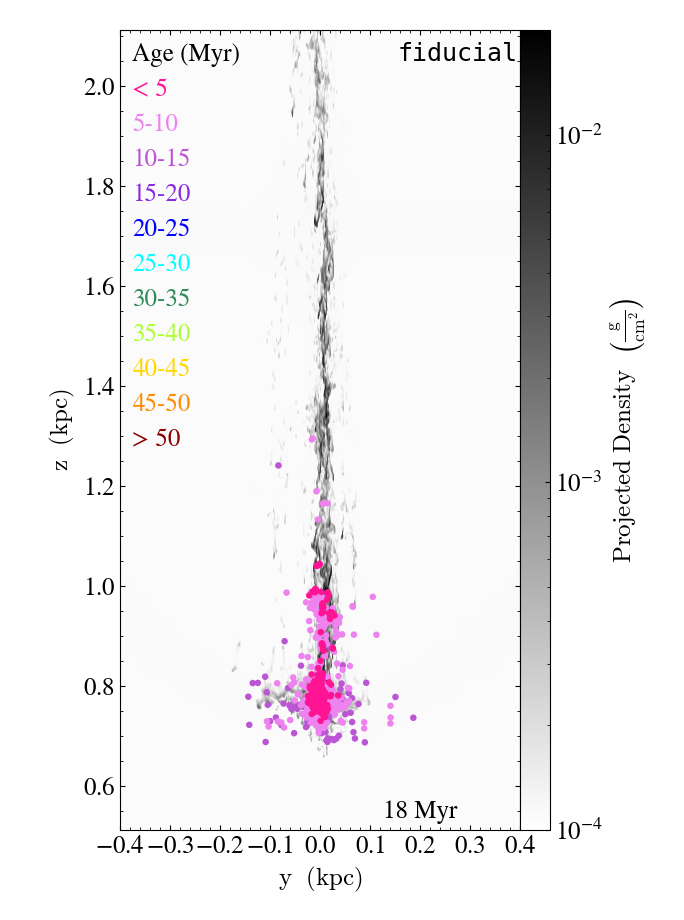}\\
    \includegraphics[scale=0.38,trim={0mm 0mm 36mm 0mm},clip]{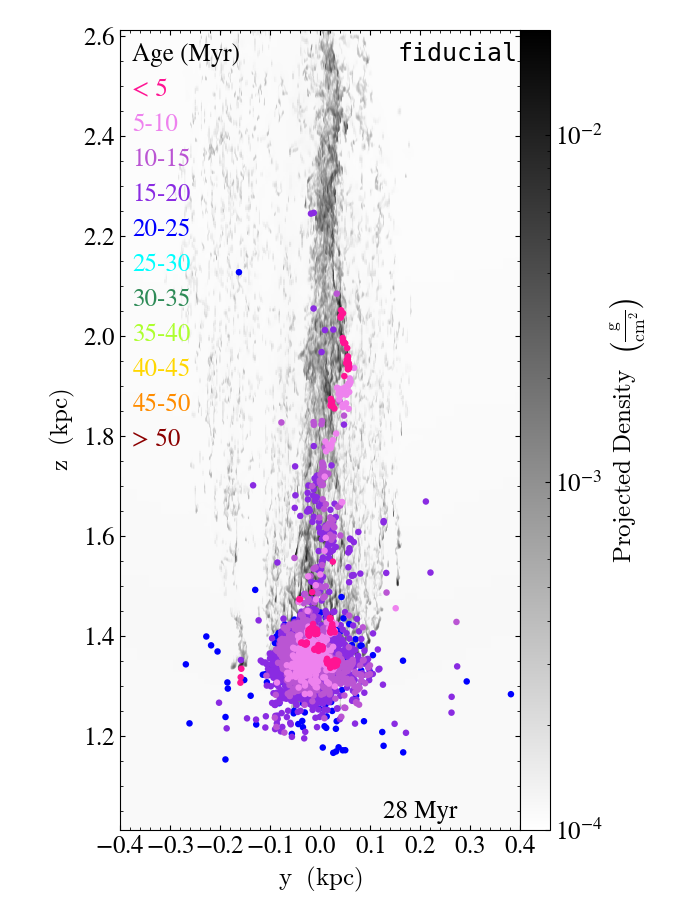}
    \includegraphics[scale=0.38,trim={0mm 0mm 36mm 0mm},clip]{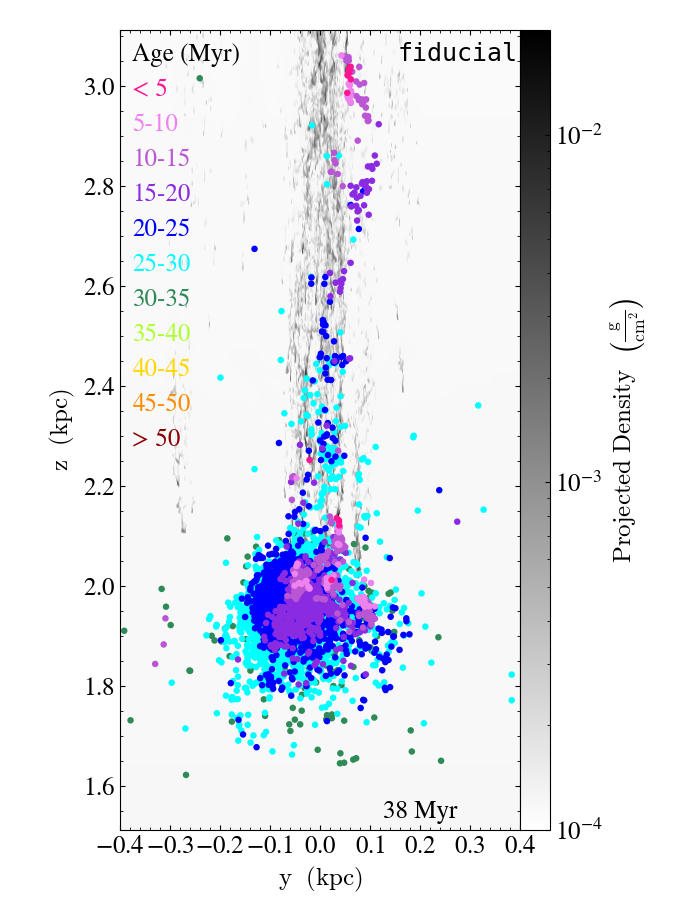}
    \includegraphics[scale=0.38]{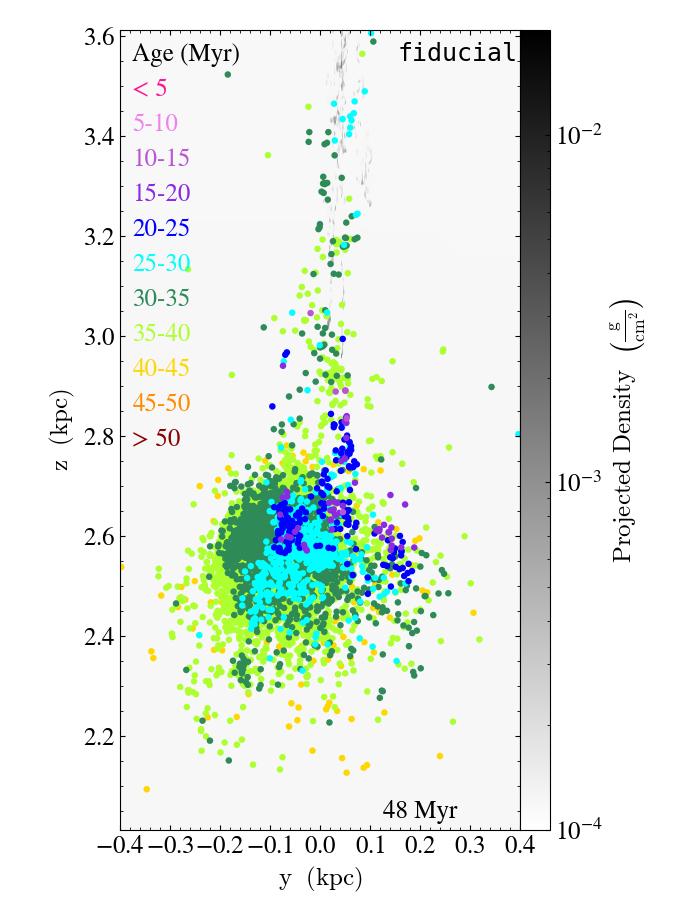}
    \caption{A time-series of the {\fid} run.  The greyscale shows the projected gas density and the points show star particles.  Particle age determines color as shown in the legends.  Particles are plotted by age such that the youngest particles are plotted on the top layer.  Each panel is 0.8 kpc by 1.6 kpc.  The horizontal axis remains centered on the same position in every projection, but the vertical axis ($z$ position) moves with the main stellar cloud.  These panels are focused on the age gradient that can be visually identified in the main star-forming clump.}
    \label{fig:projections}
\end{figure*}

\begin{figure*}
    \centering
    \includegraphics[scale=0.75,trim={11mm 0mm 36.5mm 0mm},clip]{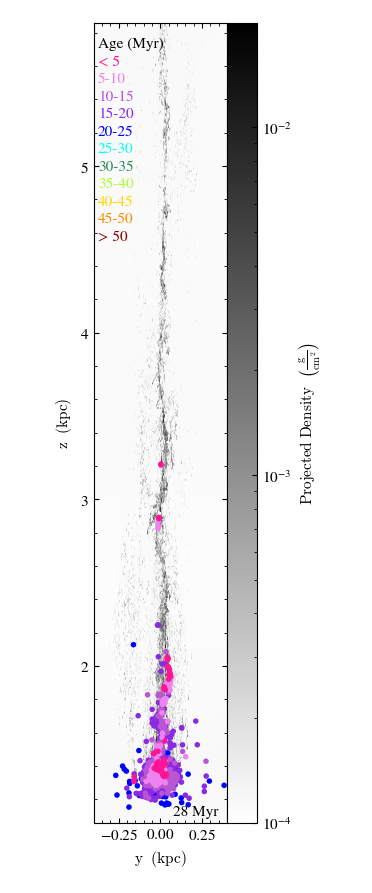}
    \includegraphics[scale=0.75,trim={24mm 0mm 36.5mm 0mm},clip]{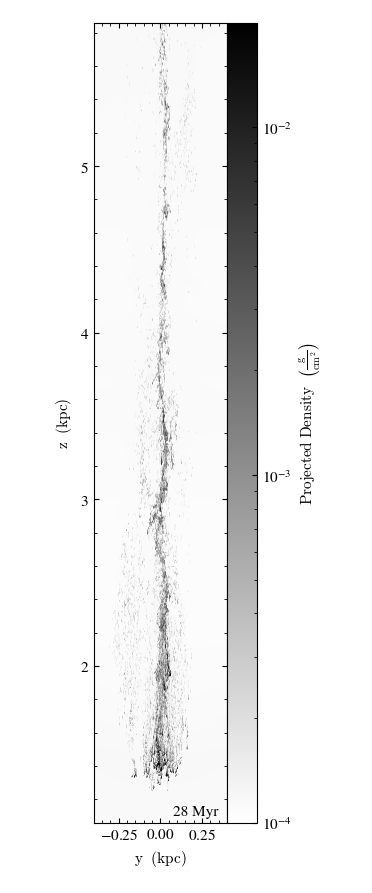}
    \includegraphics[scale=0.75,trim={11mm 0mm 36.5mm 0mm},clip]{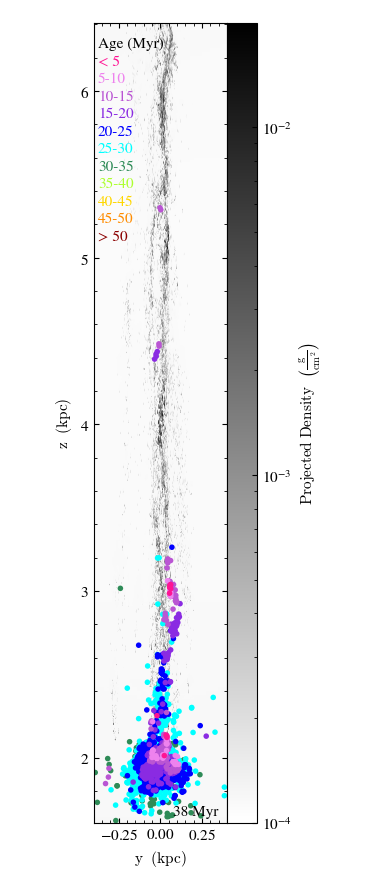}
    \includegraphics[scale=0.75,trim={24mm 0mm 36.5mm 0mm},clip]{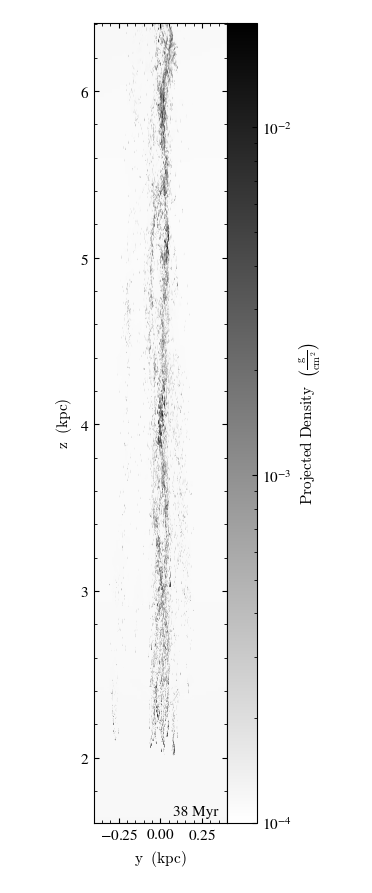}
    \includegraphics[scale=0.75,trim={11mm 0mm 12mm 0mm},clip]{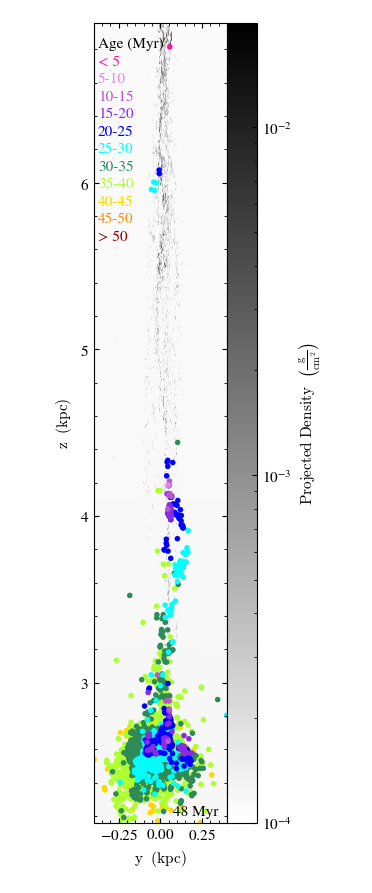}\\
    
    \caption{A time-series of the {\fid} run showing 0.8 kpc by 4.8 kpc windows of the last three outputs from Figure \ref{fig:projections}.  The greyscale shows the projected gas density and the points show star particles.  Particle age determines color as shown in the legends.  Particles are plotted by age such that the youngest particles are plotted on the top layer.  The x-axis remains centered at the same position in every projection, but the y-axis moves with the main stellar cloud.  The $z$ range of these projections has been chosen to show star particles separating from the main stellar clump as gas is being accelerated as well as stars being formed in the shredded cloud tail.}
    \label{fig:projections4p8kpc}
\end{figure*}

We begin in Figure \ref{fig:projections} with a time series of the {\fid} run, zooming in on the main stellar component (for a full view of the simulated region, please see Appendix \ref{sec:appA}).  The greyscale shows the projected gas density and the points show star particles.  Particle color is determined by stellar population age as shown in the legends.  Particles are plotted such that the youngest particles are plotted on the top layer.  The horizontal axis remains the same in every projection, but while the extent of the vertical axis ($z$ position) remains the same, its position moves with the main stellar cloud so as to keep it near the bottom of the panel.  In every panel, the (0,0) coordinates are the original $(y,z)$ cloud center position: (12.5,3.75) kpc in the simulated region.

In this figure we are focused on the region very close to the main stellar clump.  This allows us to see details of the gas cloud being stretched and shredded, and how this aligns with SF. The first panel shows the compression wave moving through the cloud before any stars have formed, illustrating that when the wind hits, the cloud has not collapsed and is still well resolved by hundreds of cells across its radius.  As the panels move forward in time we see stars forming.  In all panels until 48 Myr, we see stars that are less than 5 Myr old (magenta). Indeed, 8 Myr into the simulation all the stars are less than 5 Myr old.  Throughout the panels a nearly spherical cloud of stars is forming the main clump, and this is fully formed by 38 Myr into the simulation, at which point SF ceases in the main clump.  By 48 Myr there are no more pink ($<$ 10 Myr) stars visible in the projection.

Interestingly, stars form in the main clump continuously throughout the first 38 Myr of the simulation, rather than forming an elongated stream with a monotonic age gradient.  At 8 Myr and 18 Myr we do see elongated SF along the stretched region of high projected gas density. However, even in the small relatively spherical central clump of stars ($r$ $\sim$ 0.1-0.2 kpc) seen from 28 Myr on, we note a trend in the stellar age distribution.  First, younger stars do tend to form in the upper half of the clump starting at about 28 Myr ($z$ $\gtrsim$ 1.35 kpc at 28 Myr and $z$ $\gtrsim$ 2 kpc at 38 Myr), possibly reflecting the growing offset between the peak gas density and the peak stellar mass.  In addition, older stars show more radial spread in their distribution, even orthogonal to the wind direction.  We posit that this is because the stars inherit the velocity of the gas from which they form, and the cloud initially collapses towards the $x=y=0$ line.  This collapse is due to the rapid increase in the ICM pressure surrounding the clump, so would only be seen in clumps stripped from satellites with plunging orbits.  Any stars formed during this collapse will have higher radial velocities and larger orbits in the stellar clump.  

We also note that the main stellar clump is moving in the positive $z$-direction (seen by the changing vertical axis values), indicating that stars form from gas that has been accelerated by the surrounding medium.  

Looking above the main stellar clump, we also see stars in the narrow extended material that has been shredded from the original cloud.  As in the main clump, we can visually identify an age gradient in the extended material.  The newest stars (pink and purple) tend to be more extended than older stars.  However, even in the tail region we see stars of multiple ages at the same height (for example at about $z$=2.2 and 2.8 kpc in the 38 Myr projection).

\begin{figure*}
    \centering
    \includegraphics[scale=0.32,trim={5mm 0mm 41mm 0mm},clip]{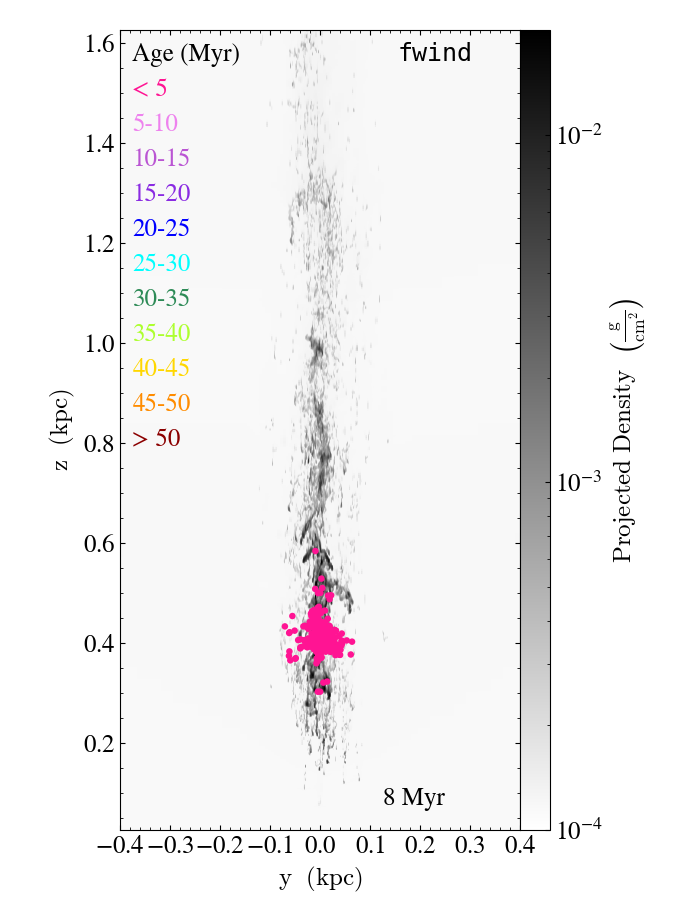}
    \includegraphics[scale=0.32,trim={5mm 0mm 41mm 0mm},clip]{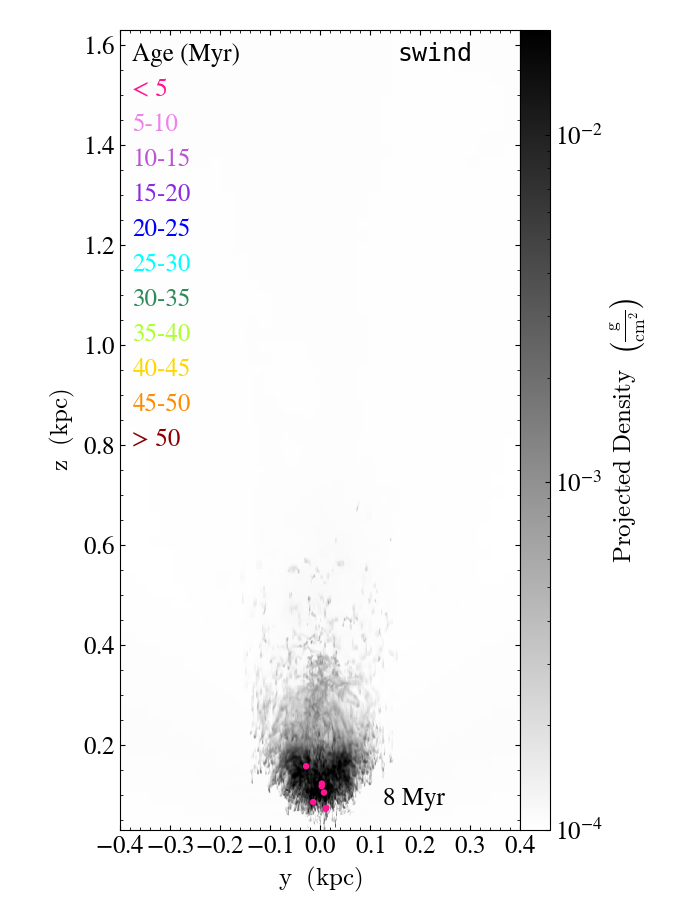}
    \includegraphics[scale=0.32,trim={5mm 0mm 41mm 0mm},clip]{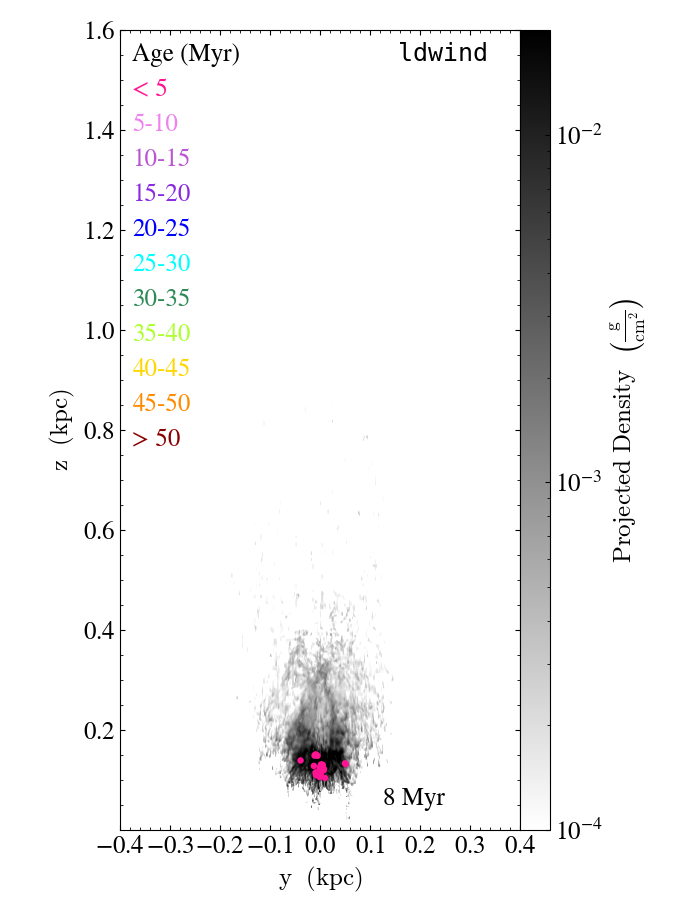}
    \includegraphics[scale=0.32,trim={5mm 0mm 0mm 0mm},clip]{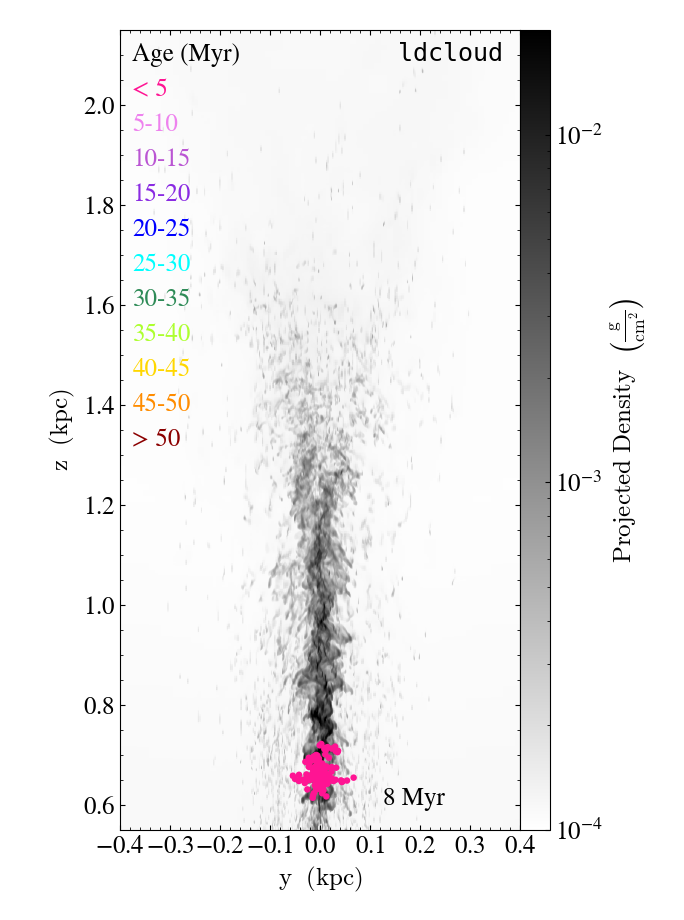}\\
    \includegraphics[scale=0.32,trim={5mm 0mm 41mm 0mm},clip]{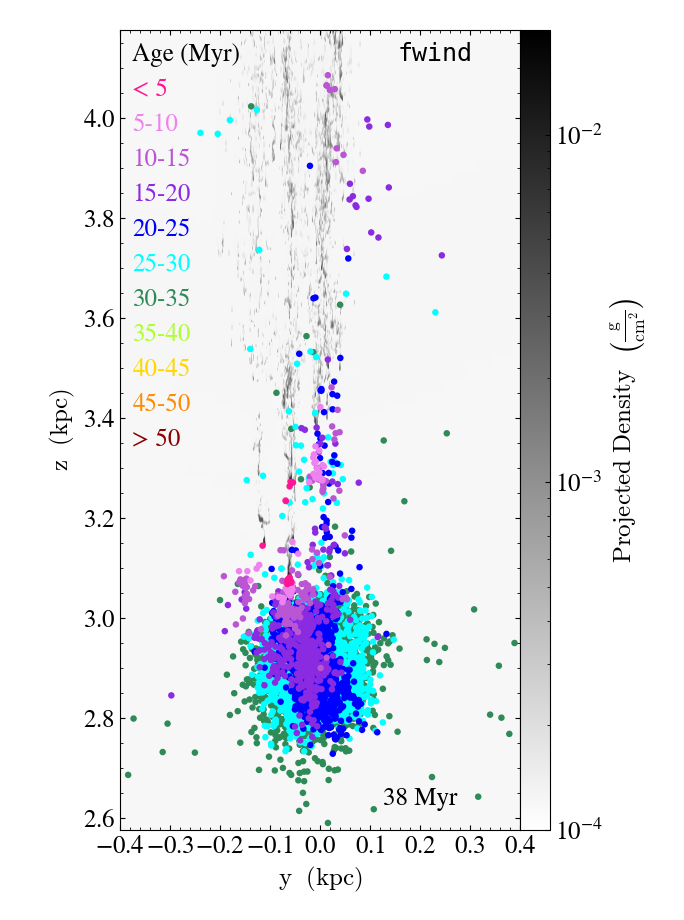}
    \includegraphics[scale=0.32,trim={5mm 0mm 41mm 0mm},clip]{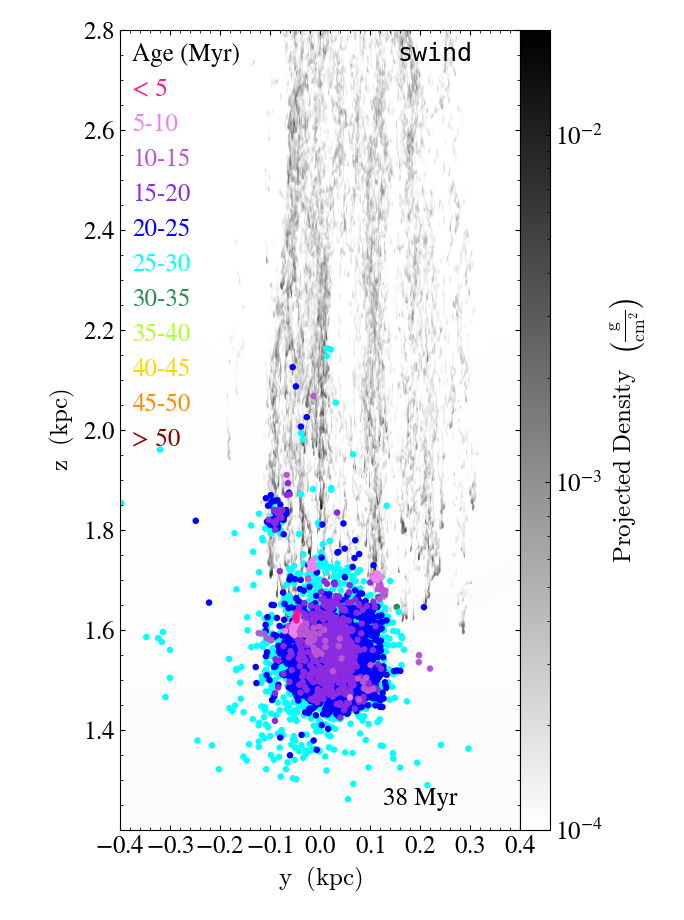}
    \includegraphics[scale=0.32,trim={5mm 0mm 41mm 0mm},clip]{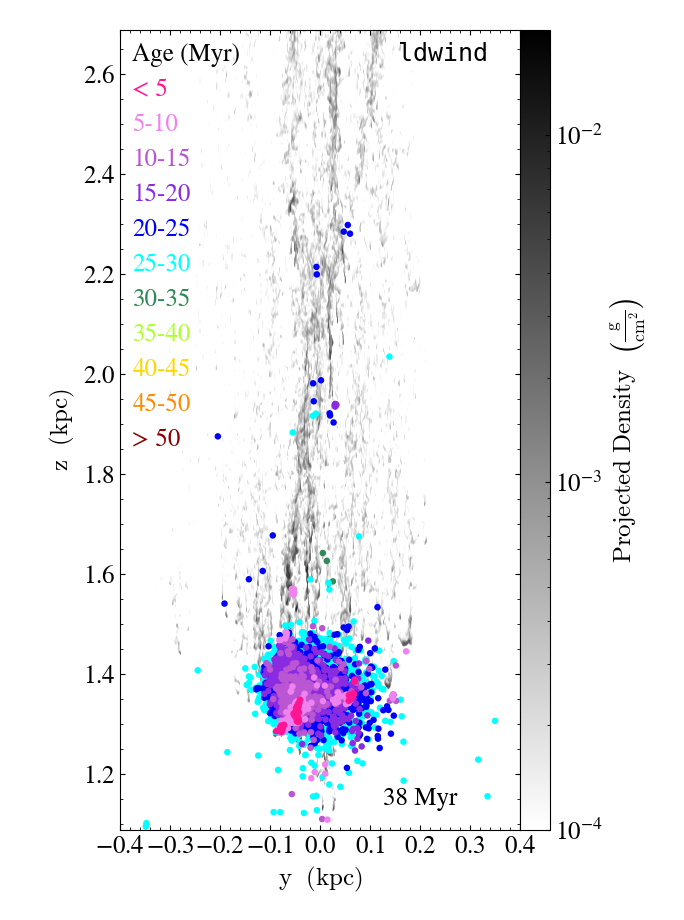}
    \includegraphics[scale=0.32,trim={5mm 0mm 0mm 0mm},clip]{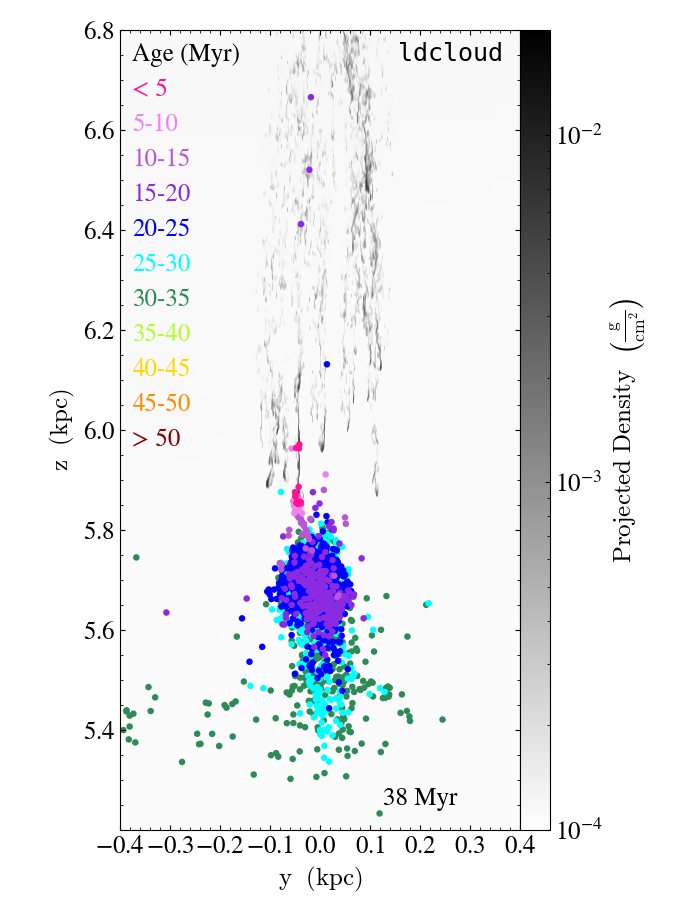}
    
    \caption{For visual comparison with the 8 Myr and 38 Myr panels in Figure \ref{fig:projections} we show the same projection in each of the comparison runs.  From left to right: {\fwind}, {\swind}, {\ldwind}, {\ldcloud}. }
    \label{fig:1p6projs_comps}
\end{figure*}

In Figure \ref{fig:projections4p8kpc} we show extended projections of the three final outputs from Figure \ref{fig:projections}.  In order to compare the stellar and gas distributions we plot only the gas density projections for the 28 Myr and 38 Myr outputs next to those with star particles.  As we would expect, the cloud gas is being accelerated by the wind and leaving the stars behind.  However, the differential displacement is small enough that at 28 Myr and 38 Myr there is still SF visually connected to the main stellar clump.  By 48 Myr, there is no dense gas below the 3 kpc $z$ position and no recent ($<$ 10 Myr) SF in the main stellar clump. 

In these projections one can see a small number of star particles forming at larger distances from the main stellar clump, highlighting that SF, particularly at later times, can occur at large (multiple kpc) distances from the main stellar clump.  Also, we note that in the projection at 48 Myr (rightmost panel), we can identify the several-kpc displacement of dense gas near the youngest star particle at $z$ $\sim$ 6.8 kpc and the main stellar clump at $z$ $\sim$ 2.6 kpc.

It is evident that the more extended downwind stellar population has a younger average age than the main stellar clump, although we will more clearly and quantitatively show this in the next section.  However, the age gradient is not monotonic even in the tail - at 48 Myr we see that the small detached upwind stellar clump at 6 kpc is older than the youngest stars in the connected extension at 4 kpc, indicating concurrent SF along several positions within the elongated gas stream. Notably, within each small star-forming clump in the extended tail there also seems to be an age gradient following the fireball scenario.

In summary, our visual impression from the time evolution of the {\fid} run is that most SF occurs in the main clump over tens of Myr, and in that main clump there is an age gradient that follows the fireball scenario of decreasing stellar age with increasing $z$ height.  However, we note that the older stars in the main clump have a more extended distribution in the z-direction than the younger stars, leading to significant overlap in the populations.  In addition, shredding, mixing, and cooling of the stripped cloud material leads to pockets of concurrent SF along a several-kpc region extending downwind of the main stellar clump. This simulation shows that the fireball scenario of a monolithic accelerated cloud forming multiple displaced stellar populations with correlated ages and heights is overly simplistic.

\subsection{Comparison of Stellar Age Distributions}\label{sec:results_compstellardist}

Now that we have visually examined the {\fid} run in detail, we will compare it with the other simulations in our suite.  

\subsubsection{Visual Comparison}\label{subsec:results_viscompstellardist}

Our first comparison focuses on the main cloud to follow the SF in the main stellar clump and to determine if the displacement of the younger particles towards higher $z$-values with respect to older stellar particles seen in the {\fid} run (Figures \ref{fig:projections} \& \ref{fig:projections4p8kpc}) can also be identified in the other runs.  In the lower panels of Figure \ref{fig:1p6projs_comps} we show the 38 Myr projections of the four comparison simulations.  The {\fid} age gradient does not seem to be a universal feature of our runs, with {\swind} showing only a possible hint of a gradient, and {\ldwind} seeming to show the opposite trend with the youngest star particles at lower $z$ values.  However, both {\fwind} and {\ldcloud} show the expected fireball gradient similar to that seen in the {\fid} run.

There are also other differences between the simulations that are visible in these projections.  First, the differences in ram pressure result in different clump positions, with the {\fwind} main stellar clump traveling farther than either the {\swind} or {\ldwind} clumps - $\sim$3 kpc from the initial cloud position versus $\sim$1.5 kpc (with {\fid} between these cases at $\sim$2 kpc).  Interestingly, {\ldcloud} has traveled the farthest of all the stellar clumps, highlighting the importance of the cloud density and radius in determining how quickly gas is accelerated and therefore how far from the parent galaxy stars will be.  Gas cloud acceleration and destruction is seen by examining the gas projection in grayscale - by 38 Myr {\fwind} and {\ldcloud} have less gas near the main stellar clump than {\swind} and {\ldwind}.  Finally, the {\fwind} and {\ldcloud} stellar clumps are elongated in the $z$-direction, while the {\swind} and {\ldwind} clouds are more spherical in shape, indicating that cloud acceleration and evolution impacts the stellar clumps morphology.  In addition, all cases except the {\ldcloud} show a similar $y$-axis extent in their stellar distribution, indicating that perhaps the initial cloud conditions determine the extent of SF perpendicular to the wind direction.

 The differences in the stellar morphology of the main clump at 38 Myr may be traced to differences in the gas cloud morphologies at early times in the cloud-wind interaction.  The upper panels of Figure \ref{fig:1p6projs_comps} show 8 Myr into the simulations, before the cloud material becomes very extended.  Interestingly, the cloud remains much more spherical in both the {\swind} and {\ldwind} runs, and has become narrower and more extended in {\fwind} and {\ldcloud}.  We also note that in {\ldcloud} one can see the outside-in removal of gas layers in the perceived flaring of the extended cloud tail.

 All in all, it seems that cases with stronger ram pressure (or lower cloud density) result in faster cloud acceleration and disruption, faster stellar clump velocities (and therefore higher $z$ positions at 38 Myr in the simulation), and a more identifiable fireball stellar morphology with older stars extending farther upwind.  In the next sections we will compare the stellar ages more quantitatively and consider stars formed farther downstream.\\

\subsubsection{Quantitative Comparison}\label{subsec:results_quantcompstellardist}

\begin{figure}
    \centering
    \includegraphics[scale=0.4,trim={5mm 12.5mm 10mm 14.5mm},clip]{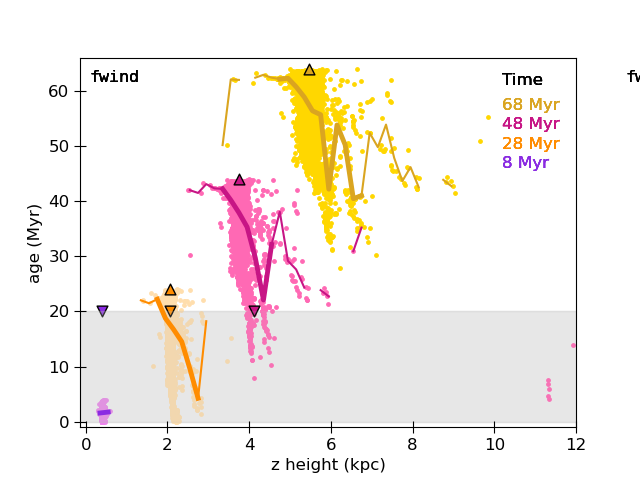}\\%
    \includegraphics[scale=0.4,trim={5mm 12.5mm 10mm 14.5mm},clip]{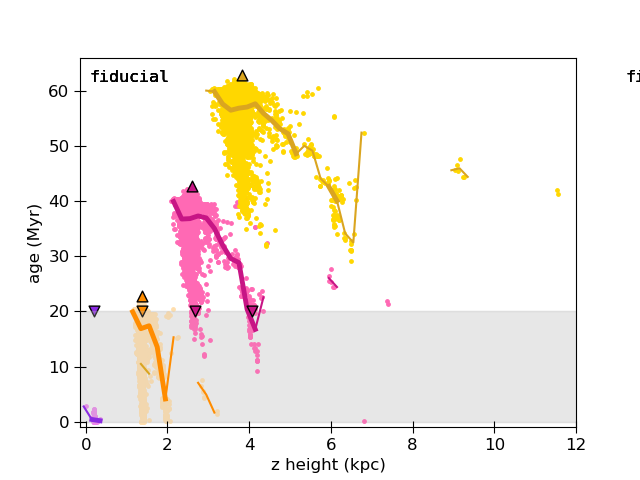}\\
    
    \includegraphics[scale=0.4,trim={5mm 12.5mm 10mm 14.5mm},clip]{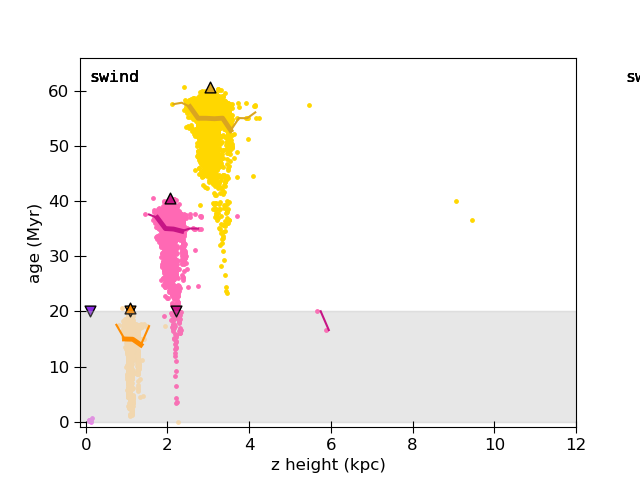}\\
    \includegraphics[scale=0.4,trim={5mm 12.5mm 10mm 14.5mm},clip]{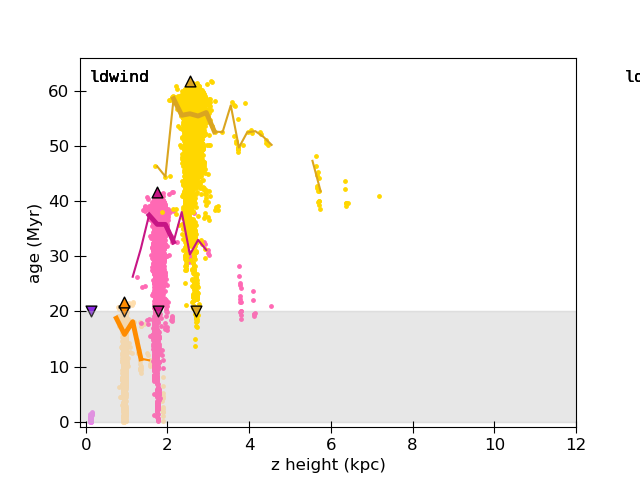}\\%
    \includegraphics[scale=0.4,trim={5mm 1mm 10mm 14.5mm},clip]{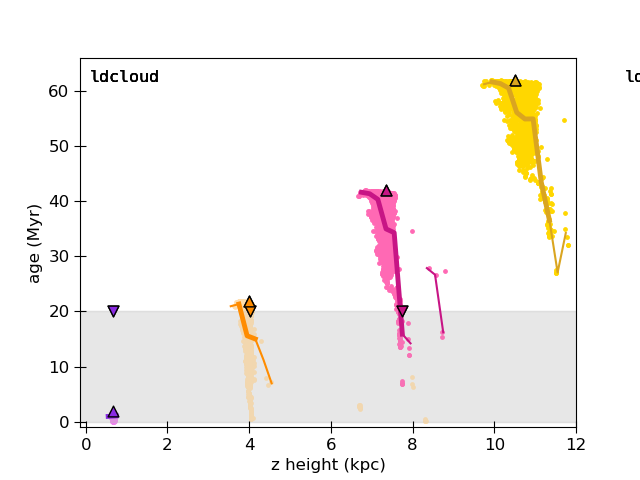}%
    \caption{Stellar age versus $z$ height at 4 times in the simulations, plotting all the stars within 12 kpc of the original position of the gas cloud.  The mean stellar age as a function of $z$ position is overplotted as a solid line, the mean position of all stars is plotted as an upward-pointing arrow, and the mean position of all stars younger than 20 Myr is plotted as a downward-pointing arrow.  All clumps show some age gradient as a function of height, although it is quite minor in both {\swind} and {\ldwind}. 
    }
    \label{fig:age_z_overplot}
\end{figure}

\begin{deluxetable}{cccc}\label{tab:clumpextent}
\tablecaption{Main Clump Extent} 
%\tablewidth{0pt}
\tablehead{\colhead{ID} &  \colhead{28 Myr} & \colhead{48 Myr} & \colhead{68 Myr} \\
\colhead{ } & \colhead{(kpc), (Myr)} & \colhead{(kpc), (Myr)} & \colhead{(kpc), (Myr)}\\
}
%\colhead{(1)} & \colhead{(2)} & \colhead{(3)} & \colhead{(4)} & \colhead{(5)} & \colhead{(6)} & \colhead{(7)} & \colhead{(8)}}\\
\startdata
\fid	&	1.0, 15.8 &	2.2,  23.2 	& 2.2, 11.3\\
 &1.2, 4.6	& 2.4, 17.4	& 4.0, 7.6\\
\ldcloud	&	0.6, 6.3     &    1.2, 25.7   & 1.6, 25.1\\
& 1.2,14.0   & 1.4, 27.4   &  2.2, 27.0\\
\ldwind	    &	0.8, 7.3 &	0.8, 7.1 &	1.4, 6.0 \\
            &	1.0, 7.6 &	2.0, 6.7$^a$ &	2.0, 3.9$^a$ \\
\swind	    &	0.6, 1.1    & 0.8, 2.4	& 1.2, 4.3 \\
            &   1.0, 0.2    & 1.4, 2.6  & 2.2, 1.5  \\
\fwind	    &	1.4, 18.0 &	1.4, 10.1	& 2.2, 21.1 \\
 &1.8, 3.8	&	3.0, 17.7 &	4.0, 19.9 \\
\enddata
\tablecomments{For each simulation and output we list the extent of the main cloud at the three later times in Figure \ref{fig:age_z_overplot} using the endpoints of the median lines in which there is at least 100 M$_{\odot}$ per 200 pc region (thick lines in Figure \ref{fig:age_z_overplot}) in the upper row, and the extent of the median lines for which there are any star particles in adjacent 200 pc regions (thin lines in Figure \ref{fig:age_z_overplot}) in the lower row.  The first value is the extent along the wind direction in kpc, and the second value is the difference in the age values at the connected endpoints.  Note that this does not always reflect the maximum age difference along the main clump.\\
$^a$ Because of scatter, the age difference at the endpoints increases along the wind direction}
\end{deluxetable}

We can examine the age gradients as a function of height more quantitatively using Figure \ref{fig:age_z_overplot}.  Each of the five panels is a different run: from top to bottom the panels are {\fwind}, {\fid}, {\swind}, {\ldwind}, and {\ldcloud}.  The colors denote different times as labeled in the legend in the top panel.  We plot the age of star particles versus their $z$-position in the domain, and we have labeled 0 on the $x$-axis as the initial gas cloud center (3.75 kpc from the bottom of the simulated region).  The lines are the median ages in bins of 200 pc in height, and if a 200 pc stretch has less than 100 M$_{\odot}$ (about 10 star particles) the median line is thin, indicating where the star particle distribution is more patchy along the tail. The differences at the endpoints of these median lines (kpc, Myr) are listed in Table \ref{tab:clumpextent}.  Note that age scatter at the edges of the clumps can lead to smaller age differences in the table than the maximum seen in the Figure. To guide the eye we have shaded ages below 20 Myr, approximately indicating stars from which we would expect H$\alpha$ emission. In addition, we note that every star particle below this line formed since the last output in this figure. Finally, for future discussion in Section \ref{sec:making_fireballs}, we have plotted upward- and downward-pointing arrows at the mean stellar $z$ height position of all stars and young stars ($<$ 20 Myr old), respectively (the $y$-axis position of the upward arrow is always at the age of the oldest stellar particle, and the downward arrow is always at 20 Myr).  

If we consider these age gradients by eye, we see similar trends to those identified in the projections.  Three of the five simulations ({\fid}, {\fwind}, and {\ldcloud}) show roughly decreasing age gradients as a function of increasing height in the fireball manner.  The other two simulations ({\swind} and {\ldwind}) show flatter age gradients, with only a hint of decreasing age with height, especially when we include the stars formed well downwind of the main stellar clump. When viewing the age distribution in these panels, however, the broad range of ages at any $z$ position within the main stellar clump becomes more clear, and we can examine the differences between our simulations and the simple fireball model in more detail.   

We will first examine the {\fid} run in detail, then discuss differences in the age distribution of stars in different runs.  First, when we focus on the median lines we see that the shapes are very similar starting at the 28 Myr output - this comes as no surprise because in all the simulations the SFR falls dramatically with time after an early peak at about 10 Myr (Figure \ref{fig:HM_SFRs}).  The 28 Myr median line shows a gradient with height - there is generally a decrease in median age with increasing $z$ throughout the main stellar clump, although there is an increase near the upper end of the clump (starting at 2 kpc).  As we can see from the distribution of points (and that the median line is thin), this increase is driven by a small number of star particles.

However, even though the median line shows a trend of decreasing age with $z$ height in line with a fireball scenario, the main clump of stars shows a 30 Myr age range at the same height.  This means that stars with the same age can be found both in the main clump and distributed along more extended filaments instead of a monotonic fireball stream.  As we will discuss in more detail in Section \ref{sec:making_fireballs}, in {\fid} there are actually two young subclumps of stars within the main stellar clump, and at the 48 Myr output both are offset from the mean position of the total stellar population.

Finally, we comment on the time evolution of the stellar distribution along the $z$ height direction ($x$-axis) in the {\fid} run.  Although the majority of stars have formed by 28 Myr into the simulation, the overall stellar stream continues to become more extended in the wind ($z$ height) direction, as can be seen in both Figure \ref{fig:age_z_overplot} and Table \ref{tab:clumpextent}.  %In the {\fid} run, the connected stellar stream composing the main clump is about 1.2 kpc long at 28 Myr, but by 68 Myr it extends about 3.8 kpc.  
Because the majority of stars have formed by 28 Myr into the simulation, the extension of the stellar clump must be largely due to velocity differences in the stars.  In the following section we will show that this morphological evolution is due to the velocity-age and velocity-height distribution of stars set at their formation.

We now compare the rest of the panels in Figure \ref{fig:age_z_overplot} to discover how the stellar age distribution is affected by different ICM winds, beginning with the similarities.  As we saw in Figure \ref{fig:HM_SFRs}, the majority of stars are formed at early times in all simulations.  We also universally find a main body of stars that shows a broad age range, indicating prolonged SF of between 30-40 Myr.  In addition, in all runs there is an expansion of the stars along the z-direction after they have been formed due to the different birth velocities of stars.  

Beyond these broad similarities, different ICM winds produce quite different stellar distributions along the wind direction. As previously seen in Figures \ref{fig:projections} \& \ref{fig:1p6projs_comps}, higher ram pressure acting on a cloud or a large lower-density cloud results in stars moving faster and therefore farther in the wind ($z$ height) direction.  

The extension of the stellar stream is also correlated with the wind speed. We define a contiguous stream to be the length of the connected median line in Figure \ref{fig:age_z_overplot}, as that means that there are no gaps of 200 pc without a single star particle. With this definition, the {\fwind} run has the longest extent of the contiguous stream.  %, going from 1.7 kpc at 28 Myr to 4.1 kpc at 68 Myr.  
On the other hand, {\swind} and {\ldwind} have much smaller contiguous main clumps at all times, only reaching a length of $\sim$2.1 kpc at 68 Myr.  %The {\ldwind} run results in a contiguous stellar stream extending $\sim$2.8 kpc at 68 Myr.  This is much shorter than the same wind speed in the {\fid} run and slightly more extended than the {\swind} run, although the region with at least 100 solar masses per 200 pc bin (thick line) is the same length in both the {\swind} and {\ldwind} runs at 48 Myr and 68 Myr. } 
Interestingly, the main stellar clump in {\ldcloud} also remains compact, and we posit that this is due to the rapid shredding and extending of the cloud material such that there are many rapidly-moving detached stellar clumps rather than a connected stellar stream (we will discuss this in more detail in the next section).  This is quite different than the reason for the compact stellar clump in {\swind} and {\ldwind}, which results from the longer survival time of a cohesive gas clump (compare the upper panels of Figure \ref{fig:1p6projs_comps}).  

Although we are able to find a gradient of lower median age with higher $z$ position in loose agreement with the fireball scenario, it is clear that the age distribution of stars is not monotonic with distance along the wind direction ($z$ position).  In fact, even the median age gradient is not always monotonic with distance as material that has been shredded and accelerated more rapidly than the main cloud condenses and forms stars (as seen in the {\fid} and {\ldwind} panels).  The scattered age distribution of stars formed from our gas clouds occurs both because the main stellar clump has a large range of stellar ages and because stars of the same age can be widely separated (several kpc).   We highlight that in {\swind} the median age gradient of the main stellar clump is nearly completely flat, indicating that with a slow wind a cloud can remain cohesive and form stars without a fireball-like age gradient.

\section{Connecting the Stellar and Gas Distributions} \label{sec:gasandstars}
\begin{figure}
    \includegraphics[scale=0.65, trim= 6.5mm 12mm 60mm 36mm, clip]{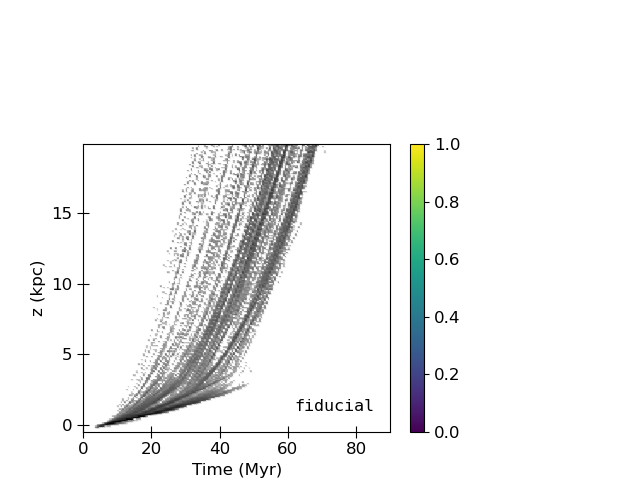}\\
    \includegraphics[scale=0.65, trim= 6.5mm 0mm 36mm 35mm, clip]{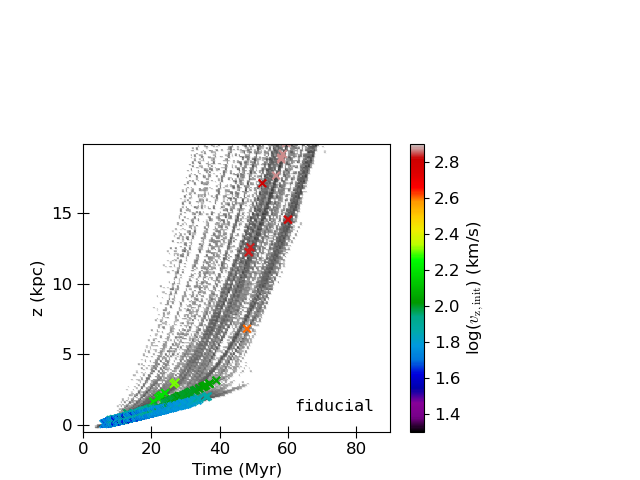}\\
    \includegraphics[scale=0.65, trim= 6.5mm 0mm 35mm 35mm, clip]{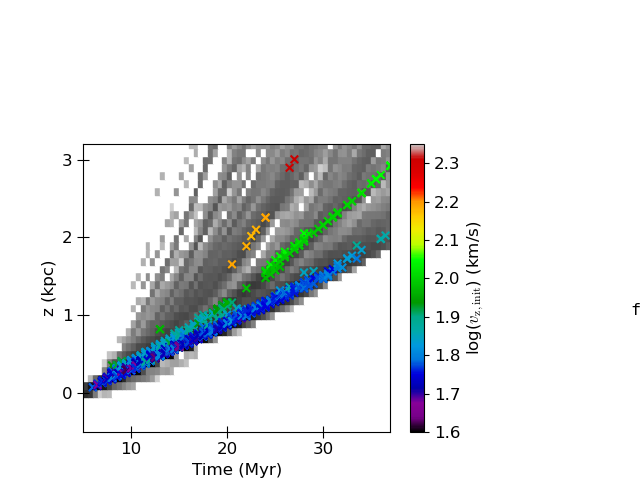}\\
    \caption{The distribution of dense ($\rho$ $\gtrsim$ 100 cm$^{-3}$) gas and newly-formed stars as a function of height and simulation time.  \textbf{Top Panel:} Grayscale 2D histograms of the dense gas distribution at every simulation output, showing significant structure in the dense gas distribution. \textbf{Middle Panel:} Star particles are plotted at their formation time and position over the dense gas distribution from the top panel.  The star particles are colored based on their initial velocity in the wind direction ($v_{\rm z,init}$). SF tends to occur in regions with more dense gas (darker gray).  At later times and large z-values stars form from gas that has been accelerated to higher velocities.  \textbf{Bottom Panel:} A zoom-in of the middle panel with a narrower $v_{\rm z,init}$ colorbar to highlight the velocity differences over time in the main stellar clump.}  
    \label{fig:zhist_fid}
\end{figure}

\begin{figure}
    \includegraphics[scale=0.65, trim= 8mm 12mm 35mm 35mm, clip]{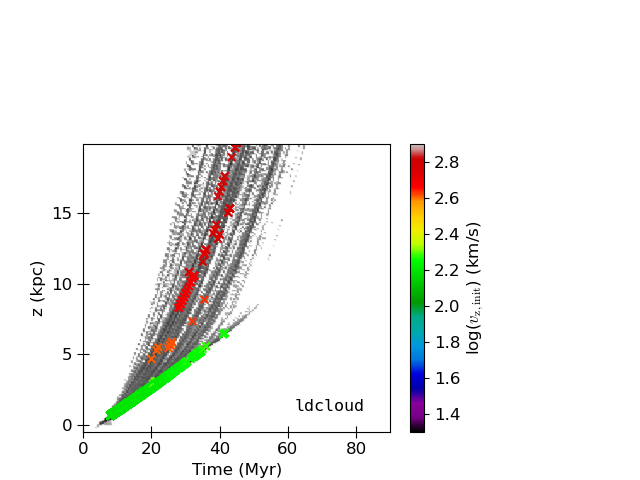}\\
    \includegraphics[scale=0.65, trim= 8mm 12mm 35mm 35mm, clip]{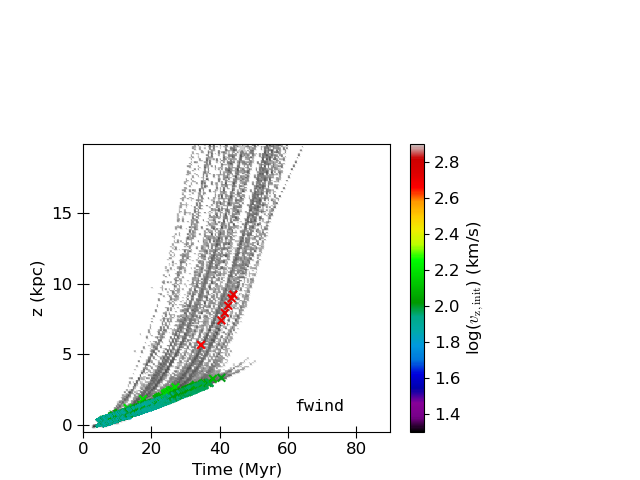}\\  
    \includegraphics[scale=0.65, trim= 8mm 12mm 35mm 35mm, clip]{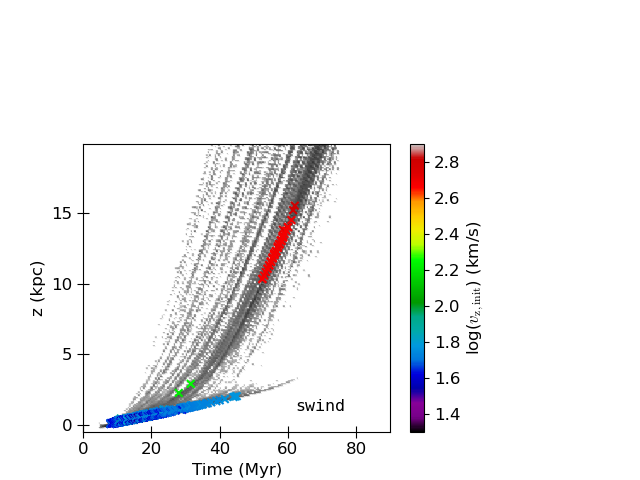}\\
    \includegraphics[scale=0.65, trim= 8mm 0mm 35mm 35mm, clip]{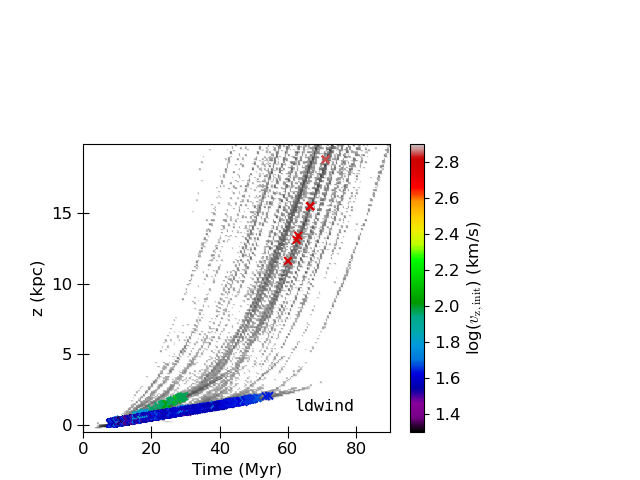}\\

    \caption{The comparison runs, with grayscale and points as in Figure \ref{fig:zhist_fid}.  In all cases the stars formed in main stellar clump, at the lowest $z$ positions, have a lower initial velocity than the stars formed from the downwind extended material.  While the velocities of the main stellar clump particles differ across the different simulations, in all runs the stars formed downwind of the main clump tend to be red, indicating similar fast formation velocities. }
    \label{fig:zhists_comp}
\end{figure}

In this section we connect the stellar age gradients, or lack thereof, to the gas distribution along the surviving main cloud and filamentary material streaming downwind of it.  

\subsection{Position and Velocity of Newly-Formed Stars}\label{sec:zhists}

To connect SF to the distribution of gas in the surviving stretched cloud, we tie the full gas distribution along the cloud to the birth positions of stars in Figures \ref{fig:zhist_fid} \& \ref{fig:zhists_comp}.  Rather than detailed images, at every output we plot a histogram of the gas mass distribution along the $z$-direction, only showing the dense gas ($\rho$ $\gtrsim$ 100 cm$^{-3}$).  The colorscale is set by the ratio of the log gas mass in the histogram cell and the log gas mass of the maximum mass in any histogram cell over the course of the simulation. We then overplot newly-formed stars as crosses such that their birth time and height are the horizontal and vertical coordinates, respectively.  Each cross is color-coded by its velocity at the time of formation (log($v_{\rm z,init}$) (km/s)).  

\subsubsection{The {\fid} Simulation}

As in previous sections, we will first discuss the {\fid} run in detail in Figure \ref{fig:zhist_fid}, then compare with our other runs (Figure \ref{fig:zhists_comp}).  If we focus on the grayscale histogram without star particles in the top panel of Figure \ref{fig:zhist_fid}, we can see the evolution of the gas distribution in the stretched cloud.  As we step forward in time and look along the vertical axis, we see that the entire dense gas component is expanding in the $z$-direction with time.  This is because gas in the cloud is accelerated at different rates - the originally more massive component at the bottom has a slowly increasing velocity (steepening slope) for the first $\sim$50 Myr until it is destroyed, while other cloud gas that has been shredded from the main cloud is accelerated more quickly.  It is also interesting to note that, while we cannot track individual parcels of gas in our grid code, when the gas distribution is viewed as a function of time as in these plots, it seems that there are identifiable curves indicating that clumps of the cloud are accelerated together. This clumping behavior is visually distinct from, for example, the uniform gray along the full length of the dense gas distribution that we would see in this histogram if the gas distribution was smooth without clumps.

When considering the range of cloud gas acceleration, we note that at any output as we look at gas higher along the $z$-direction, the slopes of the curves steepen, indicating higher velocities of gas.  This of course must be the case as the cloud material all starts within a radius of $\sim$200 pc, but by 40 Myr has extended to about 20 kpc.  In addition, if we follow any of the gas curves that peel off the original cloud body, we see that over the course of tens of Myr and/or $\sim$5 kpc the curve rapidly steepens and then seems to settle at a higher velocity (which is still roughly a factor of two below the ICM velocity, in agreement with \citet{Tonnesen_2010_multiphasetails} that dense gas does not reach the ICM velocity).  

In the second panel in Figure \ref{fig:zhist_fid} we overplot the newly-formed stars.  Each star particle is plotted once on this panel, at the time and $z$ position of its creation.  While the majority of stars (more than 99\% of stars) are formed within the most slowly-moving clump at the bottom of the panel, we see that SF does occur in some of the stretched and fragmented cloud material. These SF episodes in the extended dense material show higher birth velocities when they form farther from the main stellar component.  We also note that when multiple crosses denoting new stars lie along a dense filament in this figure, these are stars that are likely forming from a single surviving dense cloud. Thus, from a small 200 pc gas cloud we find stars extending nearly 20 kpc and showing an overall age gradient along that length.  Despite this population-level gradient, as we have shown in Figure \ref{fig:age_z_overplot} and discuss further below, the formation time of stellar particles does not monotonically increase with height.

Finally, in the bottom panel we zoom-in to early times, low $z$-values, and a constrained velocity colorbar to illustrate two main points about the velocity differences in newly-formed stars.  First, tracking a single chain of formed stars forward in time or up in $z$ position (height in the wind direction) results in higher stellar velocities.  This is what we expect from the fireball scenario in which a cloud of gas accelerates and forms stars such that those formed early are formed at lower $z$ positions and slower velocities.  The gas cloud accelerates and leaves these stars behind, and at later times forms stars at higher $z$ positions and faster velocities.  This acceleration produces an initial age gradient, and the difference in stellar velocities will result in the $z$ expansion of the stellar distributions at later times that we see in Figure \ref{fig:age_z_overplot}.  On the other hand and opposing the formation of the monotonic fireball stellar age distribution, stars can form at multiple heights along the stretched cloud at the same time (there are crosses at three clearly separated heights at various times between 20-30 Myr).  Stars formed at the same time but at different positions along the stretched cloud are formed with different velocities - higher velocities for stars formed farther along the cloud.  This means that stars of the same age can be found at wide separations throughout a stretched cloud.     

We have already quantitatively shown there is not a monotonic stellar age gradient in the main clump in Figure \ref{fig:age_z_overplot}.  Putting the dense gas distribution together with the newly-formed stars, we find that the long filamentary distribution of dense gas in the extended cold cloud further complicates the toy fireball scenario. Because the cloud stream as a whole forms coeval stars at multiple heights, the full stellar component formed from our original cloud does not form a single fireball.  However, stars formed in the condensed cold clumps within the larger-scale stream do tend to reproduce the age-height and velocity-height gradient expected from the fireball scenario.  Therefore, multiple stellar clumps separated by several kpcs within a stripped tail could be forming from a single massive gas cloud being shredded such that the cold material is nonuniformly accelerated away from the galaxy.

\subsubsection{Varying the Wind Parameters}

We can compare the {\fid} run to the other simulations by comparing Figures \ref{fig:zhist_fid} and \ref{fig:zhists_comp}.  In an overall view, all of the clouds evolve in a similar fashion: a massive component of the cloud is accelerated slowly in which the bulk of the stars form (the ``main clump''), the extent of the gas cloud expands as it is continually shredded and cooled in a long clumpy tail, and a few stars form in the extended clumpy material at late times.  

However, cloud evolution differs based on the wind and cloud parameters.  The three bottom panels in Figure \ref{fig:zhists_comp} show the gas distribution over time with newly formed stars overplotted for the varying wind parameter runs.  The main clump clearly survives for a longer time in the {\swind} and {\ldwind} runs than in the {\fid} and {\fwind} runs, and forms stars for a longer time.  However, only in the {\ldwind} does this result in a global increase in the mass of stars formed in the simulation (Table \ref{tab:clouds}).  We note that the main clump survives and forms stars for about the same amount of time in both the {\fid} and {\fwind} runs. This may be because the amount of dense gas that survives in the main clump is quite similar in both runs, even though their overall morphology shows a narrower cloud in the {\fwind} run (compare Figures \ref{fig:projections} \& \ref{fig:1p6projs_comps}). %

Looking at the extended material, we see that dense gas remains in the box the longest for {\ldwind}, then {\swind}, {\fid}, and finally {\fwind}.  Even though the cloud crushing time is initially nearly identical for the clouds in {\ldwind} and {\swind}, the drag time is shorter for the {\swind} cloud\footnote{$t_{\rm drag}$ $\sim$ $\chi$ $r_{\rm cloud}$ / $v_{\rm wind}$ while $t_{\rm cc}$ $\sim$ $\chi^{1/2}$ $r_{\rm cloud}$ / $v_{wind}$, where $\chi$ is the density ratio of the cloud and wind}.  This may help explain why cloud material is seen in the box for longer in {\ldwind} (and why the main clump is at lower $z$-values in {\ldwind} than in {\swind} at the same time as seen in Figure \ref{fig:1p6projs_comps}). We also note that in the {\ldwind} run dense material continues to be removed from the main clump at later times than in the {\swind} run, although most of this gas does not form stars.  

The SF in this extended material traces the regions that have the most dense gas, or the darkest shaded regions. However, even in the {\swind} run, which has the most SF in the extended region of the varying wind runs, only 0.5\% of the total stellar mass is formed beyond the main clump.  

In summary, in all wind cases we see continuous SF in the ``main'' clump until just before the main cloud of dense gas is exhausted or diffused into its surroundings.  In each simulation, stars formed later in this main clump tend to have higher velocities than stars formed earlier (best seen in the zoom-in of the {\fid} $z$-position versus time panel in Figure \ref{fig:zhist_fid}), meaning that there is an age-height and age-velocity correlation at star formation.
 Second, we see that in all cases SF is not confined to the main clump, and occurs in clumps that have separated and been accelerated to higher velocities by the wind.  In simulations with the same ICM gas density ({\fwind}, {\fid}, and {\swind}), more SF occurs in the extended material within slower winds that do not quickly destroy small dense clumps.  We also find that a low-density wind ({\ldwind}) may not result in significant mass in the extended clumpy dense gas, leading to less SF far from the main clump.   
 
\subsubsection{Varying the Cloud Density}

We see that the cloud properties also effect the cloud evolution by focusing on the top panel in Figure \ref{fig:zhists_comp}.  The {\ldcloud} run, with a lower density and larger cloud, shows stronger acceleration in the main clump, resulting in faster formation velocities for the stars in the main clump.  The rapid acceleration of cloud gas also results in all the dense cloud material leaving the simulated region earlier than in {\fid}.  Interestingly, more gas is distributed throughout the extended cloud (the gray is darker across more of the histogram than in other runs) in {\ldcloud}.  This is likely because the extended gas distribution of this cloud means the outer regions are more weakly gravitationally bound and can therefore be accelerated from the main clump more quickly than the smaller {\fid} cloud material, allowing for rapid mixing with the ICM and momentum transfer \citep{tonnesen_RPSclouds_2021}. The rapid shredding of the original main cloud allows for a large amount of dense gas in the extended cloud tail that forms the most star particles of any run, although it is still only 3.5\% of the total amount of stars formed in {\ldcloud}.   The extended SF beyond the main clump occurs in several small components that can be visually seen as connected chains of newly-formed stars in this panel.

Briefly, a larger and lower-density cloud with the same mass as {\fid} has less SF overall due to less SF in the rapidly shredded main clump, but more SF distributed throughout the extended clumpy material of the filamentary cloud. The stars formed in the main clump of {\ldcloud} have higher velocities than those in {\fid} because of the rapid acceleration of the main gas cloud.

\subsection{Velocity Evolution of Stars}\label{sec:stellarvelocityevolution}
\begin{figure}
    \centering
    \includegraphics[scale = 0.65, trim=3mm 0mm 20mm 20mm, clip]{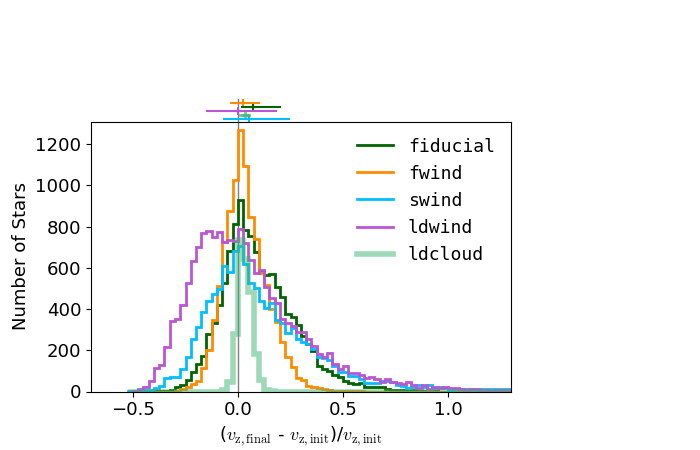}\\
    \includegraphics[scale = 0.65, trim=3mm 0mm 20mm 20mm, clip]{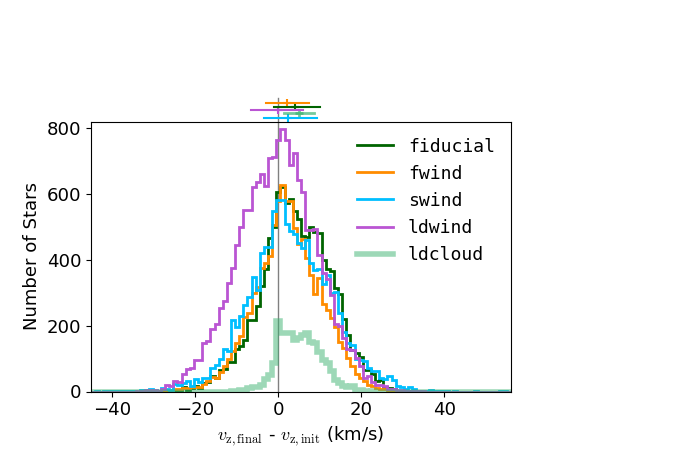}
    \caption{Histograms of the ratio of the velocity change, measured as the difference between the particle velocity at the final output at 100 Myr ($v_{\rm z,final}$) and the initial (formation) velocity of the particle ($v_{\rm z,init}$), to the initial velocity of star particles (top panel) and the measured velocity change of star particles (bottom panel) in the simulations as labeled.  The gray vertical line is at zero change, and above each panel we plot the median and interquartile range for each run (the lines are offset vertically for clarity).  The velocity of most stars shows little change over the course of any of the runs, with a tendency for the velocity to increase in the wind direction (positive median values).  The width of the distribution in both the upper and lower panels increases for the runs with lower ram pressure ({\swind} and {\ldwind}), whose dense gas clouds stay coherent for a longer time. }
    \label{fig:vzratio_hist}
\end{figure}

\begin{figure}
    \centering
    
    \includegraphics[scale = 0.56, trim=4mm 11.5mm 20mm 36mm, clip]{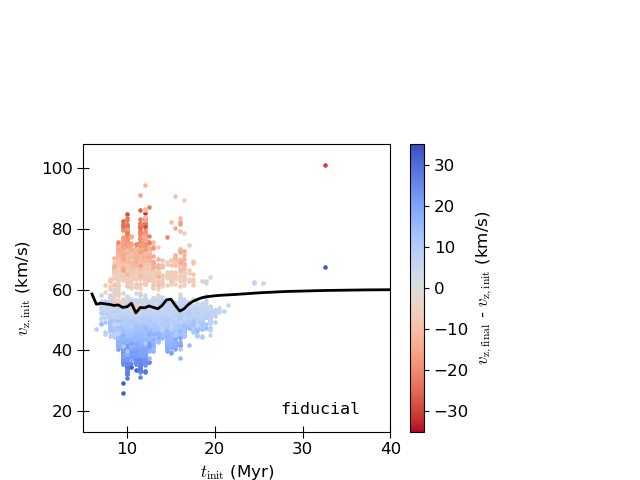}\\
    
    \includegraphics[scale = 0.56, trim=4mm 11.5mm 20mm 36mm, clip]{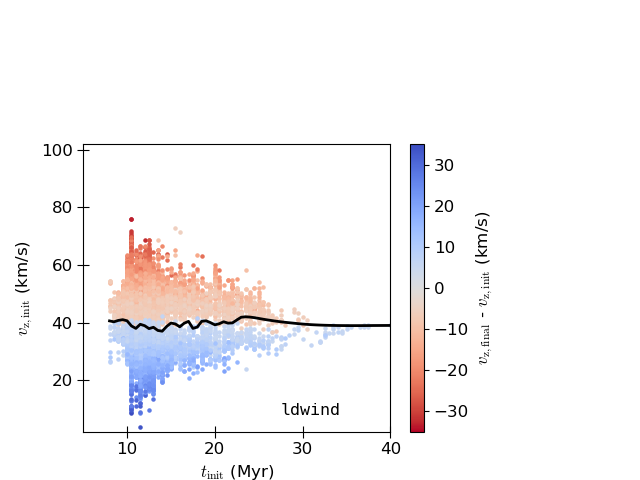}\\
    \includegraphics[scale = 0.56, trim=4mm 0mm 20mm 36mm, clip]{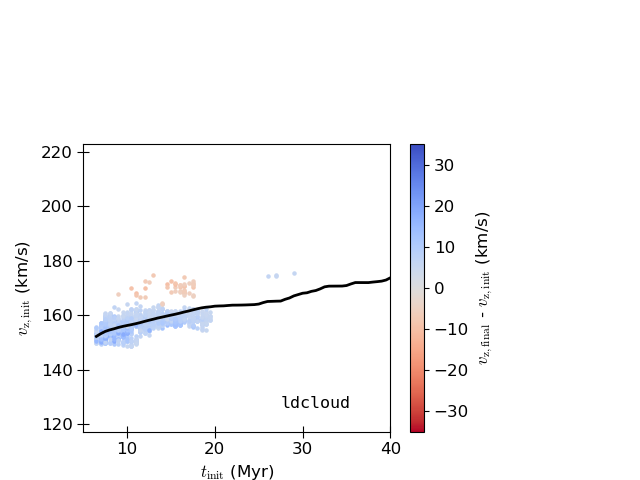}\\
    \caption{$v_{\rm z, init}$ versus $t_{\rm init}$ for star particles whose velocity changes by more than 5 km/s, color-coded by their velocity change.  The solid black line is the mass-weighted average velocity of the full stellar population.  Stars faster than the average tend to slow down, while those slower tend to speed up, particularly for the {\fid} and {\ldcloud} simulations.  The gravity from the accelerating dense cloud of gas is likely responsible for the blue (accelerating) star particles that are moving faster than the average stellar velocity.}
    \label{fig:vzinit_tinit_vzdiff_scatter}
\end{figure}

In Figures \ref{fig:zhist_fid} and \ref{fig:zhists_comp} we found that at formation, stars are found near high density gas with increasing velocity as they are formed at higher $z$ (higher positions along the wind direction) in the simulated region.  Because the star particles do not interact hydrodynamically with the surrounding wind there may be an implicit assumption that their velocities will remain constant, and the combination of the velocity-height and age-height gradients found in stars at their formation (Figures \ref{fig:zhist_fid} \& \ref{fig:zhists_comp}) will result in long-lived stellar gradients that roughly follow the fireball scenario.  However, gravitational acceleration either from the gas or from the other stars could change stellar velocities, either enhancing or depressing any initial velocity differences and therefore possibly affecting any initial age gradient in the stellar population.  In this section we measure whether and to what extent the velocity of stars changes after formation. 

In Figure \ref{fig:vzratio_hist} we plot histograms of the relative and absolute change in velocity of star particles after formation, ($v_{\rm z,final}$ - $v_{\rm z,init}$)/ $v_{\rm z,init}$ and ($v_{\rm z,final}$ - $v_{\rm z,init}$), in the top and bottom panels, respectively.  This change is measured over the course of the entire simulation, such that $v_{\rm z,init}$ is taken at each particle's formation time, and $v_{\rm z,final}$ is taken at last snapshot of the simulations, in other words at 100 Myr.  Note that positive values indicate that the velocity increased after star particle formation.  In all simulations, the peak of the distribution is around zero, indicating that there is generally not a significant change in the velocities of stars after formation.  

However, we do see differences between the simulations.  In the lower ram pressure {\swind} and {\ldwind} runs, the distribution of ($v_{\rm z,final}$ - $v_{\rm z,init}$)/ $v_{\rm z,init}$ is larger (seen both in the overall histograms and in the interquartile ranges drawn above), showing both more star particles slowing after formation and a tail of particles that increase their velocities by a large fraction.  Some of this is due to the low initial velocities of star particles in these simulations, as seen in Figure \ref{fig:zhists_comp}, which means that a small change in velocity could result in a larger ratio.  However, upon inspection of the lower panel, we also find that the absolute velocity differences also span a larger range particularly for the {\ldwind} run, although there is a slightly larger distribution of ($v_{\rm z,final}$ - $v_{\rm z,init}$) in {\swind} as well.  As we move to higher ram pressure ({\fwind}) or lower cloud density ({\ldcloud}), $v_{\rm z,final}$ tends to be closer to $v_{\rm z,init}$.

While Figure \ref{fig:vzratio_hist} gives an overall picture of how the velocity of star particles change in our simulations, we need to dig deeper to see which stars are changing their velocities and why.  For that we turn to Figure \ref{fig:vzinit_tinit_vzdiff_scatter}, which shows the $v_{\rm z,init}$ as a function of formation time ($t_{\rm init}$) for all star particles for which the absolute value of ($v_{\rm z,final}$ - $v_{\rm z,init}$) is greater than 5 km s$^{-1}$.  Each point is color-coded by the absolute change in $v_{\rm z}$ such that red points slow down and blue points speed up. We have also overplotted the mass-weighted mean velocity of all the stars across the simulations as a black solid line.  We show this distribution for three simulations: {\fid}, {\ldcloud}, and {\ldwind} from top to bottom.% 

We immediately see that the $\Delta$$v_{\rm z}$ of star particles acts to decrease velocity differences in the vertical stratification of ($v_{\rm z,final}$ - $v_{\rm z,init}$) of star particles across all three simulations:  those with faster $v_{\rm z,init}$ values slow down, and those with slower $v_{\rm z,init}$ values speed up. As we look at simulations in which the cloud is accelerated more rapidly and has more filamentary mass downstream (Figures \ref{fig:zhist_fid} \& \ref{fig:zhists_comp}), the $v_{\rm z,init}$ value at which stars begin to slow down moves farther above the average stellar velocity.  This is likely because the gas cloud has much more mass than the stellar clump at all times, so in simulations in which it is rapidly accelerated from the stellar clump it will tend to pull stars to higher velocities until it is far enough away that the gravitational acceleration is minimal.  

It might be initially surprising that $\Delta$$v_{\rm z}$ shows a strong relation with the $v_{\rm z,init}$ offset rather than one with the offset in $z_{\rm init}$ relative to the stellar clump position (not shown).  However, stars in the main clump form close to the center of mass of the clump, well within $r_{\rm clump}$ $<$ 100 pc in the $z$ direction (see Figures \ref{fig:projections} \& \ref{fig:1p6projs_comps}). Therefore, even if a star forms below the stellar center of mass, a velocity offset of 10 km/s will result in crossing the center of mass within a few Myr, at which point the star would be slowed down by the stellar clump.  Perhaps if gas clumps were large enough that stars moving faster (slower) than the average clump velocity could not cross the clump in a few Myr we would see more of a trend with $z_{\rm init}$ as well as $v_{\rm z,init}$.

In Figure \ref{fig:vzinit_tinit_vzdiff_scatter} the largest changes in $v_{\rm z}$ tend to occur in star particles that are formed earlier in the simulations.  We argue that this is related to the survival of the main gas clump in two main ways.  This is first because gravity will have the largest effect when there is a large amount of mass nearby, whether that is the main dense gas clump at early times or the main stellar clump at later times.  The gas clump is about ten times the mass of the main stellar clump, and will therefore more strongly draw star particles until it is shredded and stretched across several kpc.  We see that stars have larger $v_{\rm z}$ changes at early times, when the gas cloud is more likely to still be coherent.  In fact, in {\ldwind} the gas cloud stays near the main stellar clump for the longest time, and star particles with large changes in their velocities (more than 5 km/s) are seen for the longest time.  On the other hand, the gas cloud in {\ldcloud} is rapidly accelerated and shredded, and stars have smaller $\Delta$$v_{\rm z}$.  The survival of the main gas cloud is also important because at early times the cloud is collapsing as well as accelerating, and therefore $v_{\rm z,init}$ can have larger offsets from the average stellar velocity.  These particles then require larger $\Delta$$v_{\rm z}$ in order to match the average stellar velocity.   %

 Therefore, we find that self-gravity can reduce the velocity distribution of stars, which will result in more compact stellar distributions and, importantly, should act to reduce any age gradient.  Notably, the largest velocity changes tend to occur relatively early in the simulations and for the low ram pressure winds, indicating that a massive coherent main gas cloud has the most effect on whether the stellar particle velocities change after they are formed.  Although we find that self-gravity affects some of the stars in all of our simulations, we point out that our stellar distributions still expand with time (Figures \ref{fig:projections} - \ref{fig:age_z_overplot}).  Self-gravity does not lead to shrinking clouds, it just slows expansion.

\section{Discussion}\label{sec:discussion}

Our simulations have given a physical underpinning to what sets the stellar distribution in cold clouds that are being accelerated by an ICM-like hot wind.  They confirm that, as described in the empirical fireball picture, individual clouds do accelerate and and leave newly-formed stars in their path.  However, in detail we find that because clouds are stretched by nonuniform acceleration and mixing with the surroundings, the position and velocity of these stars is intricately determined by the cloud-wind interaction.  In this section we will first place our simulations in the context of previous work studying cloud evolution using simulations, use our results to gain insight into what is required to make a fireball-like age gradient, and then discuss how these simulations compare to observations.  We will end the section with a discussion of the limitations of this study.%

\subsection{The Impact of Gas Cloud Evolution on the Stellar Distribution}\label{sec:discussion_gasclouds}

As we discuss in the Introduction, several simulations of cold clouds interacting with hot winds have found that short radiative cooling timescales can result in the growth of cold clouds rather than their destruction \citep{GronkeOh2018_coldcloudgrowth,GronkeOh2020a_coldgasentrains,Armillotta2016_diskcorona,Marinacci2010_coldgasaccretion,Abruzzo_2022ApJ...925..199A}.  As pointed out by \citet{GronkeOh2018_coldcloudgrowth}, in order to observe the eventual growth of clouds, simulations must follow them for a long period of time (several cloud-crushing times) and a correspondingly long distance (tens to hundreds of cloud radii).  Because of this feature, simulations are able to follow the evolution of the originally spherical cloud as it becomes shredded into long filamentary structures, and find that it is within these filamentary structures that the mixing-driven cooling of the surrounding material occurs.  

Our simulations also find extended filaments of dense gas (Figure \ref{fig:projections4p8kpc}), and that dense gas can not only survive the entire length of our simulation domain but form stars several to tens of kpc above the main star-forming clump (Figure \ref{fig:zhists_comp}).  The survival of cold gas throughout the full simulation box is no surprise when we compare the cloud crushing time to the cooling time of mixed material.  The cloud crushing time, $t_{\rm cc} = \chi^{1/2} r_{\rm cloud}/v_{\rm wind}$, extends from about 2 Myr ({\fwind}) to 4 Myr ({\swind} and {\ldwind}).  However, the cloud itself rapidly cools to about 10 K ($\sim$0.05 Myr, see Section \ref{sec:SFR}), so the temperature of the mixing layer is about 10$^4$ K, near the peak of the cooling curve.  Even at the wind density of 10$^{-26}$ g cm$^{-3}$, at solar metallicity the cooling times of the mixing layer are less than 2 Myr.  Thus in all of our cases we are in the growth regime as defined in \citet{GronkeOh2018_coldcloudgrowth}.%The SF downwind of the main clump within the filaments muddies any height-age gradient because the filamentary SF can be concurrent with SF in the main clump.
 
Given that cooling can add dense material to the extended filaments streaming from an otherwise destroyed clump, we might expect significant SF in these filaments.  However, quite generally we find that the majority of our stars form in a main clump.  We argue that this behavior may be largely due to the rapid onset of SF in our simulations.  In our {\fid} simulations, the wind first interacts with the cloud about 2.5 Myr into the simulation and our first stars form less than 3.5 Myr later, which is about a half of the cloud-crushing time. About 90\% of the stars formed throughout the entire simulation have formed within 12 Myr from the initial wind interaction (about 2 cloud-crushing times).  Cloud-crushing simulations without SF find that within the first few to several cloud-crushing times of the wind interaction, the amount of dense gas decreases before beginning to increase through mixing-layer cooling.  However, in most simulations our SF happens fast enough that we can form spherical stellar clumps before much of the original cloud has been heated and stretched into filaments.  We do find that our larger cloud, {\ldcloud}, is rapidly accelerated, so even though some stars form in the first few Myr, the gas cloud rapidly stretches so that the stellar distribution is more elongated.

However, although most of our SF occurs in the main cloud, we note that only about 10\% of the original cloud mass forms stars in the wind cases, and the rest of the gas flows downwind forming clumps and filaments (Figures \ref{fig:projections4p8kpc}, \ref{fig:zhist_fid}, \& \ref{fig:zhists_comp}).  As we will discuss more in Section \ref{sec:disc_caveats}, rigorously simulating the sparse SF in the filamentary downwind material would require SF recipes carefully calibrated for the small amount of mass in any cell.  Based on this work we conclude that stars can form in the dense gas downwind of the main cloud, but save our careful comparisons with observations (Section \ref{sec:disc_observations}) for the main clumps that form thousands of star particles.

In future work, delaying SF through non-thermal pressure support in the cloud or by changing the SF criteria (discussed more in Section \ref{sec:disc_caveats}) would be an interesting test to determine if more SF would occur in the extended filamentary cloud if the main clump was shredded before significant SF occurred within it or if a massive stellar clump can coexist with significant SF downstream.

\subsection{How to Make a Fireball}\label{sec:making_fireballs}

In the majority of our simulations we do not strictly reproduce the empirical fireball scenario, as illustrated in Figure \ref{fig:fireball_illustration}.  However, the age gradient found in the main stellar clump of {\ldcloud} quite closely follows the fireball pattern, perhaps most easily seen in Figure \ref{fig:age_z_overplot}.  Thus, here we focus on what sets {\ldcloud} apart from the other simulations.  The {\ldcloud} cloud begins with a larger radius and a lower density than any other of the simulations, and this then results in a more rapid acceleration of the cloud (compare the $z$ distribution along the vertical axis in Figure \ref{fig:1p6projs_comps} or the $z$ distribution of dense gas and newly formed stars in Figure \ref{fig:zhists_comp} for the different runs).  Because the cloud has a high acceleration, the formation time gradient of stars as a function of $z$ position in {\ldcloud} is steeper than in any other simulation.  This is most clearly seen in Figure \ref{fig:zhists_comp}.  Therefore, barring much post-formation velocity evolution in the star particles we would expect to see a fireball age gradient, and indeed, {\ldcloud} has the least velocity evolution in the star particles (Figures \ref{fig:vzratio_hist} \& \ref{fig:vzinit_tinit_vzdiff_scatter}).   

It seems then that in order for a star-forming cloud to produce a fireball morphology, it needs to be forming stars as it is rapidly accelerating, and the mass distribution needs to be extended enough that the stars do not lose their initial differential velocity via the cloud's self-gravity.  This storyline agrees with the trends we see in Figure \ref{fig:age_z_overplot}, in which the average age of the stars decreases with height in {\fwind} and in {\fid} as well as {\ldcloud}, but not in the lower ram pressure cases.  However, unlike {\ldcloud}, {\fid} (and also to a less extent {\fwind}) has a large distribution in ages seen at the base of the main stellar clump, which we posit is because the initial cooling and collapse of the denser cloud allows a small high-density cloud to survive and undergo little acceleration via interaction with the wind.  

The main feature of the {\fwind} age distribution in the $z$ direction that does not follow the fireball model is the jumpy distribution of the average age of stars as a function of height in Figure \ref{fig:age_z_overplot}.  This behavior is less intuitive, but we argue that it is due to the scatter in the initial velocities of the star particles as the cloud both collapses and accelerates.  Note that at 28 Myr, there is a relatively clear age gradient with height, with only a few particles at about 3 kpc that are older than expected.  If the predicted velocity gradient that younger stars should have higher velocities than older stars held, this age gradient should have remained smooth over time and become more apparent.  Instead some of the older stars have high velocities, making the age gradient become less smooth over time.  

This highlights a key factor in creating a clear fireball age gradient: the main clump must be accelerated fast enough that the overall relation between age, height, and velocity of the stars cannot be overshadowed by scatter in formation velocities or by nearby downstream SF from the collapse of the filamentary structures.  The extremely rapid acceleration of the {\ldcloud} makes any small formation velocity differences of stars unimportant to the maintenance of the overall age-height gradient.  Further, the rapid acceleration of the extended filamentary material means that stars formed downwind of the main clump are separated by a few to several kpcs from the more massive main stellar clump and seem to be separate stellar streams.

We can consider this in terms of a few timescales.  First, how quickly the cloud is accelerated depends on how rapidly momentum is deposited into the material via mixing.  While there has been excellent work on the growth time of clouds by, for example, \citet{Tan_2023_Cloudywithrain}, here we use the simple assumptions presented in \citet{tonnesen_RPSclouds_2021} that clouds are spherical, remain the same size, and accrete all the wind that would pass the cloud cross-section.  That wind then changes the velocity of the cloud while it is increasing the fraction of gas in the cloud originating from the wind.  We can rewrite equation (7) in that paper as:
\begin{equation}\label{eqn:tfICM}
    t_{\rm fICM} = \frac{f_{\rm ICM}}{1-f_{\rm ICM}} (\frac{4}{3}) \chi \frac{r_{\rm cloud}}{v_{\rm wind}}
\end{equation} 

so that the time to reach a chosen ICM fraction ($f_{\rm ICM}$), or equivalently the time to reach the same fraction of the wind velocity, is a factor of $(\frac{4}{3}) \frac{f_{\rm ICM}}{1-f_{\rm ICM}}$ times $t_{\rm drag}$.  Using our cloud initial conditions, the time it takes for our clouds to reach 100 km s$^{\rm -1}$ (or 0.1 $f_{\rm ICM}$ for our $v_{\rm wind}$ = 1000 km s$^{\rm -1}$ runs) is lowest for {\ldcloud}, at only 1.3 Myr.  The first stars form about 3 Myr after the wind hits, at which point the cloud has already accelerated to a high velocity (seen in Figure \ref{fig:vzinit_tinit_vzdiff_scatter}).  The next shortest time to reach 100 km s$^{\rm -1}$ according to the above equation is {\fwind}, at 2.8 Myr.  However, {\fwind} forms stars about 2 Myr after the wind hits the cloud, before the cloud has reached a high velocity.  This may be why the scatter in the formation velocities in {\fwind} is able to result in a nonmonotonic age gradient.  The rest of the simulations take even longer to reach high velocities. %$\sim$ $\chi$ $r_{cloud}$ / $v_{wind}$ while t$_{cc}$ $\sim$ $\chi^{1/2}$ $r_{cloud}$ / $v_{wind}$, where $\chi$ is the density ratio of the cloud and wind}

Although we do not vary our cloud parameters beyond the {\ldcloud} run, with this intuitive understanding we can briefly discuss what we might expect from changing our cloud mass.  First, we can increase our cloud mass by increasing the cloud density or size.  From Equation \ref{eqn:tfICM} we can see that increasing the mass by increasing the density will result in a directly proportional increase in $t_{\rm fICM}$, while increasing the volume will only increase the $t_{\rm fICM}$ by mass to the 1/3 power.  Therefore, in order to have a more massive cloud accelerate fast enough to produce a dramatic fireball gradient it likely has to have a larger radius than {\ldcloud}.  On the other hand, lowering the mass of the gas cloud by either lowering the density or by shrinking the cloud could still allow for a strong age gradient along the tail, at least until the cloud is small enough that the cloud crushing time is smaller than the cooling time and it would therefore dissipate into the wind.

Notice that in this discussion we have focused on the main stellar clump in considering whether we reproduce the classic fireball scenario.  There is a straightforward reason for this: although there is quite extended SF that by construction in our simulation is originating from a single small cloud, in observations it would likely be extremely difficult to convincingly prove that small clumps of sometimes coeval forming stars separated by more than a kpc (and surrounded by many other unconnected star-forming clumps within the stripped tail) are in fact connected by filamentary gas originating from the same gas cloud.  Therefore, although we can find an age gradient when smoothing over the long ($\sim$20 kpc) filamentary structure formed from our single cloud, we do not believe that this is currently an observationally relevant prediction.  

\subsection{These Simulations in an Observational Context}\label{sec:disc_observations}

Here we first compare our simulations to observations of individual stellar clumps, and then discuss if we can use these simulations to make predictions for the stellar distributions across entire jellyfish tails. 

This subsection focuses on comparisons with the GASP sample \citep{poggianti_gaspI_2017, Poggianti_2025A&A_GASPsummarypaper}, because they have the largest sample of ram pressure stripped galaxies with star formation in their tails.  Not only was this a large survey of ram pressure stripped galaxies, but the observed sample was selected from B band images of cluster galaxies, likely predisposing it to have a large fraction of star formation in the stripped tails.  Indeed, in \citet{Gullieuszik_2020_GASP_SFRsintails}, of the 54 ram pressure stripped galaxies examined, only 17 had SFRs in their stripped tails of less than 10$^{-2}$ M$_{\odot}$ yr${-1}$.  Other surveys, such as the VESTIGE survey of the Virgo cluster, have found very few tails with star formation in the ram pressure stripped population \citep{Boselli_2018_VESTIGEintro, Boselli_VESTIGE_SFtail_2018A&A...615A.114B}.  Using the GASP sample, \citet{Gullieuszik_2020_GASP_SFRsintails} finds that while there is not a single parameter determining whether SF will occur in a stripped tail, overall, more tail SF is found from massive galaxies that are stripped near the center of low-mass clusters, reasoning that a significant amount of dense gas needs to be stripped in order to form stars.

There are also a series of GASP papers that study the stellar clumps in detail \citep{poggianti_gaspSFoutsidegalaxies_2019, Giunchi_2023_cloudmorphology, Giunchi_2023ApJ...949...72G, Werle_SFtails_2024A&A...682A.162W,Gullieuszik_SFclumpsGASP_2023ApJ...945...54G}, with \citet{Giunchi_2023_cloudmorphology} examining the morphology of more than 300 H$\alpha$ and/or UV star-forming clumps, as well as more than 400 optical complexes in the stripped tails of six galaxies.  We compare our results with this work now.

\subsubsection{Comparison with individual star forming Complexes}

Throughout this work, we have compared our simulations to the empirically-developed fireball scenario with a monotonic age gradient, finding our stellar distributions to be more complicated.  We now more carefully compare our stellar clumps to observations and find overall agreement.  

Although many observations show evidence for an age-height gradient as predicted by the fireball model \citep[e.g.][]{Kenney_2014_fireballs, Smith_2010_Comafireballs,Jachym_2017_D100,Giunchi_2023_cloudmorphology,Werle_SFtails_2024A&A...682A.162W}, when observing a sample of more than 300 optical star-forming clumps that have H$\alpha$ emission, \citet{Giunchi_2023_cloudmorphology} find that only a small majority ($\sim$60\%) of young star-forming complexes show the expected fireball age gradient.  Therefore we argue that it is not a huge surprise that we rarely see a clean fireball gradient in our small suite of simulations.  

We also highlight that \citet{Giunchi_2023_cloudmorphology} determines the age gradient of star forming regions using the displacement of the centers of the H$\alpha$, UV, and optical clumps, measured as the intensity-weighted mean position of the clump in the plane of the sky.  In the cases shown in that work, even when there is a displacement between the center of the H$\alpha$ and optical emission, the H$\alpha$ emitting region is embedded in the optical star-forming complex, with optical emission on all sides.  Allowing for the younger stars to be embedded in a more extended older population, and only requiring a displacement of the mean position of the young stars results in all of our main star forming clumps showing the fireball age displacement at some point in their evolution.  We illustrate this in Figure \ref{fig:age_z_overplot} by plotting triangle symbols at the mean position of the stars within a few kpc of the main star-forming clump.  At each time the downward-pointing triangles show the mean $z$ position of all stars below 20 Myr in age, roughly denoting an H$\alpha$-emitting population (and always plotted at 20 Myr on the $y$-axis), and the upward-pointing arrows show the mean position of all stars near the main clump (always plotted at the age of the oldest star particle on the $y
$-axis).  In all of our simulations there is an offset between the total population and the young population mean positions at the 48 Myr snapshot (shown in pink).  As a point of interest, we note that in the {\fid} run, within the main stellar clump there are two well-separated stellar regions with stars less than 20 Myr old in the 48 Myr snapshot.  Such a configuration has indeed been observed; for example, in \citet{Giunchi_2023_cloudmorphology}, about 4\% of star forming complexes contain more than one H$\alpha$ emitting clump.

Finally, we note that although the total amount of SF is low in the filamentary structures downwind of the main stellar clump, dense star-forming gas is accelerating quite rapidly.  Therefore, the trails of stars formed within these filaments do show fireball-type age gradients.  We are referring to the streams of visually connected red (high-velocity) `x' markers in Figure \ref{fig:zhists_comp}, most clearly seen in the {\ldcloud} and {\swind} panels.  We would expect these age gradients to form because the dense gas in the filaments is moving even more rapidly than the {\ldcloud} clump, and as we explained in Section \ref{sec:making_fireballs}, accelerating star-forming gas will allow for a stellar age gradient along the cloud path.  

Therefore, although we do not see a simple scatter-free stellar age gradient in our simulations, when we more carefully compare our results to observations, particularly the large clump dataset in \citet{Giunchi_2023_cloudmorphology} \citep[see also][]{Giunchi_2023ApJ...949...72G,Werle_SFtails_2024A&A...682A.162W}, we find that our results are not in tension with observations.  Instead, we physically explain the variety of clump morphologies observed using our suite of simulations, which can be summarized as: rapid clump acceleration and destruction will be the most likely to form extended stellar clumps in the wind direction with well-separated young and old populations.  Indeed, \citet{Werle_SFtails_2024A&A...682A.162W} do find that elongated stellar clumps tend to have larger displacements between the H$\alpha$ clump and the center of the star-forming complex.  We can go a step further and predict that the longer stellar clumps will be highly displaced from any molecular gas clouds or the gas clouds will be completely shredded due to the strong ram pressure.

A more comprehensive parameter study would be required to reproduce the observed population of star-forming clouds in ram pressure stripped tails.  While this is beyond the scope of this work, our simulations indicate that varying cloud properties to larger and lower density clouds could result in more dramatic fireball morphologies, perhaps even at lower at the lower ram pressure experienced in cluster outskirts and groups.  In addition, a focus on finding the wind and cloud parameters for which the main gas cloud survives and forms stars for hundreds of Myr would lead to a better understanding of the various formation pathways of the overall observed stellar clump population. % 

\subsubsection{Clouds as a Component of Full Jellyfish Tails}

These simulations are of individual clouds interacting with an ICM wind, but they represent a small component of a much more massive jellyfish tail that has been stripped from a satellite galaxy.  We refer to this larger component as `the complete jellyfish tail'. Here we comment on how being part of this much larger system may effect the clouds' evolution and whether we can use our simulations to understand stellar distributions on the larger scale of a complete jellyfish tail.  

First we will discuss how the tail environment might affect cloud evolution.  Our simulations include the self-gravity of the cloud system, but do not include the gravitational potential of the galaxy from which the gas has been stripped.  Simple tests show that clouds of our mass and an even larger initial radius ($\sim$1 kpc) will stretch by less than 15\% over 50 Myr, even when their initial position is only 1 kpc from the disk plane of a galaxy with a Milky-Way like mass.  At larger distances from the disk plane the stretching is reduced further. Therefore, the gravitational potential of a cloud's host galaxy is not likely to have a strong impact on its morphology.

The velocity structure within the complete jellyfish tail could also impact cloud evolution and survival.  We choose a fiducial wind velocity near our small clouds of 1000 km s$^{-1}$, but not only could this vary dramatically within a tail, measuring the 3D velocity of a cluster satellite galaxy is rarely possible (see \citet{Rasmussen_2006_RPStail_XrayMachspeed, Ignesti_2023A&A_RPSradiocontinuum}). We note that even the line of sight velocities of several jellyfish galaxies within the GASP survey are more than 1000 km s$^{-1}$ \citep{Gullieuszik_2020_GASP_SFRsintails}. 3D models of jellyfish galaxies with star formation indicate high velocities, such as more than 3000 km s$^{-1}$ for ESO 137-001 \citep{Vollmer_2024_ESO137model}.  Within the stripped tail, simulations have found that dense gas tends to have velocities below the overall wind velocity, ranging to well over 1000 km s$^{-1}$ slower than the wind \citep{Tonnesen_2010_multiphasetails, tonnesen_RPSclouds_2021}.  Because the mixing layer between a cold dense cloud and the hot ICM gas may have a range of velocities as well as densities and temperatures, we measured the velocity of hot gas within 1 kpc of the center of the clumps identified in \citet{tonnesen_RPSclouds_2021}, and found a large range of gas velocities near the clumps.  For the 1000 km s$^{-1}$ winds, the distribution of the velocity near the clumps peaked at about 900 km s$^{-1}$ throughout the tail, while in the 3230 km s$^{-1}$ wind, the velocity near the clumps tended to be 1500 km s$^{-1}$ at 15 kpc above the disk and rose to 2500 km s$^{-1}$ near the edge of the box.  Based on these studies, while a relative velocity of 1000 km s$^{-1}$ seems reasonable, a suite including a larger range of relative velocities is warranted.

In addition, in our simulations the wind is laminar, but observations indicate that ram pressure stripped tails can be turbulent \citep[e.g.][]{Li_2023_ESO137_tailturbulence,Ignesti_2024_GASPtails_turbulence_VSF}.  Simulations have shown that turbulence in the surrounding wind can enhance cloud destruction, particularly in smaller clouds or in those with cooling timescales close to their cloud destruction timescales \citep{Gronke_2022_cloudcrushing_turbulence, VergisGeorge_2025arXiv_cloudcrushing_turbulence}.  Because our clouds are large and rapidly cool, ICM turbulence may not have a large impact on the survival of our main clump, but future work should test this.

When applying our results to observations of complete jellyfish tails, we must not only focus on how a stripped cloud evolves in a given wind, but must also consider the properties of clouds that will be stripped by the ICM interaction.  For example, in this work we find that a faster wind will result in less cloud SF than a slower wind.  However, we cannot claim that tails of radially-plunging galaxies will have less overall SF than those with slower orbital velocities.  This is because one expects more total gas to be stripped from galaxies by fast winds as well as the removal of denser, more massive clouds that could form more stars.  Our small-scale results are not in conflict, therefore, with observational results that indicate that there is more SF in tails affected by strong ram pressure \citep[e.g.][]{jaffe_gaspphasespace_2018,Gullieuszik_2020_GASP_SFRsintails}.

We finally want to highlight that our single-cloud simulations cannot be easily translated into predictions for the age distribution of stars throughout a complete jellyfish tail.  Stripped jellyfish tails that have SF tend to have several tens of star-forming complexes, and young and old stellar populations are distributed over a wide range of distances from the host galaxy \citep[e.g.][]{Werle_SFtails_2024A&A...682A.162W}.  This has been studied in detail in galaxy-scale simulations by \citet{Akerman_SFtails_2025A&A...698A.151A}, who find that across a tail stars form at different heights and velocities.  This is because gas at a range of densities can be stripped concurrently from a galaxy disk, and therefore the stripped gas will accelerate from the galaxy and collapse to form stars at varying rates \citep[see also differential stripping and acceleration discussed in][]{Tonnesen_2010_multiphasetails, tonnesen_RPSclouds_2021}.  Because gas stripping is a continuous process, within the complete jellyfish tail it is observationally very difficult to disentangle when any individual cloud was stripped from the disk, how fast it was accelerated, and how quickly it began to form stars. \citet{Akerman_SFtails_2025A&A...698A.151A} also clearly shows that projection effects can muddy observational attempts to separate stars into individual trails.  

Therefore, our idealized wind-tunnel simulations can be used to understand how individual clouds would evolve within a complete jellyfish tail.  However, we cannot use our simulations to predict stellar distributions across complete jellyfish tails because the cloud population will differ according to galaxy and ICM properties.

\subsection{Simulation Limitations}\label{sec:disc_caveats}

Our simulations have allowed us to gain a physical understanding of what causes and disrupts a fireball age gradient in a ram pressure stripped gas cloud.  However, we are unable to run a test suite of simulations varying all possible subgrid models such as those used in our SF prescriptions, nor do we include all of the physical (micro-)processes that are likely acting in this environment (such as magnetic fields, viscosity, thermal conduction, and turbulent pressure support).  In this section we will discuss how changing our physical prescriptions could affect our results.  

First, although as discussed in Section \ref{sec:method} we chose reasonable parameters for our SF threshold and efficiency in comparison to other works \citep[e.g.][]{Marinacci_SMUGGLE_2019MNRAS.489.4233M}, and as we mention in Section \ref{sec:SFR} the final stellar masses of our clumps are within the observed clump mass ranges in \citet{Giunchi_2023_cloudmorphology, Werle_SFtails_2024A&A...682A.162W,Giunchi_2025_GASP_clump_mass_function}, we do test some variations on our parameters.  We find that changing our SF density threshold from 100 cm$^{-3}$ to 1500 cm$^{-3}$ (to match the Jeans Mass of most of our dense gas) does little to vary the SFR in our {\fid} simulation.  In addition to our increased density threshold for SF, we change the efficiency to 10\% from 2\%, and find that early SF is slightly enhanced and later SF is somewhat depressed.  We can next ask if we can use the gas and stellar mass of clumps to constrain the star formation efficiency we should apply in our simulations.  The stellar mass of observed clumps ranges from below 10$^4$ M$_{\rm \odot}$ to above 10$^7$ M$_{\odot}$ \citep{Giunchi_2025_GASP_clump_mass_function}, and because we cannot observationally connect the original stripped gas cloud masses directly to the stellar mass of the clumps they will form in the tail, we cannot use the stellar mass formed from the gas clouds in our simulations to constrain our star formation recipe.  However, the lack of observational constraints does not affect our conclusions, which are based on comparisons between our simulation suite.  While we can change the details of the SF history by changing the parameters of our subgrid model, this does not change the main insights of our paper into what causes the fireball age gradient.    

We also did not test changes to our stellar feedback prescription.  In our simulations we can visually find evidence of supernovae about 10 Myr after star formation in the sudden blowout of gas and expansion of the dense stream near and downwind of the stars (compare the gas tail width at 18 Myr and 28 Myr in Figure \ref{fig:projections}), but it could be interesting to see whether strong early feedback from stellar radiation and/or winds would slow overall star formation and allow for more cloud acceleration and subsequently more extended stellar distributions.

While we find that our results on the stellar distribution of the main cloud (which in all runs is composed of thousands of star particles) are robust, the SF distribution within the filamentary material downwind of the main clump may be more influenced by our specific SF recipe.  We find material at our SF threshold density in the filaments, but how much of that gas would truly collapse to form stars could vary with our SF efficiency or minimum stellar mass.  Simulations that form individual stars down to very low masses and that could simulate SF on a cloud basis rather than a cell basis would be extremely informative in understanding the details of filamentary downwind SF.  Thus, in this paper, while we have discussed downwind SF, we have not quantitatively compared it between simulations in detail.

We chose a simple initial condition for our clouds, that of being in pressure equilibrium with a pre-wind surrounding material.  The cloud cools rapidly and isochorically, such that once the higher-pressure ICM wind hits the cloud it can be compressed to fulfill out star forming criteria.  In future work, testing lower density clouds that will cool more slowly and testing clouds that have realistic nonthermal pressure support (turbulent and magnetic) that could delay cloud collapse and SF \citep[e.g.][]{Appel_2022ApJ_Bfields_and_turbulence_on_SFR} would allow us to make predictions for how long we might expect clouds to survive within the ICM wind before forming stars.  In addition, \citet{BandaBarragan_2018_cloudcrushing_cloudturbulence} found that turbulence within the cold cloud itself can result in more rapid acceleration and impact the downwind gas structure.  If a more turbulent cloud accelerates more rapidly, it could therefore show stronger age gradients in the wind direction, as long as the cloud collapsed enough for stars to form during the rapid acceleration phase.  Determining the interplay of rapid acceleration and delayed collapse that could be caused by including magnetic fields and cloud turbulence in forming fireball-like age gradients would require a dedicated suite of simulations.  In this work we instead focus on what happens once cooling and collapse has driven the cloud to begin forming stars.

These are purely hydrodynamic simulations, and therefore we do not include magnetic fields in either the cloud or the surrounding wind.  Some work has shown that ICM magnetic fields could drape over clouds, producing magnetic drag that couples the cloud to the surrounding gas and results in more rapid acceleration \citep{Dursi_2007ApJ_magneticdraping2D_noKH, Dursi_Pfrommer_2008ApJ_magneticdraping, McCourt_2015_magnetizedclouds}. The magnetic tension from these magnetic field lines should also suppress the Kelvin-Helmholtz (KH) instability, as seen in KH simulations \citep[e.g.][]{Jones_1997ApJ_bfield_noKH,Lyutikov_2006MNRAS_magneticdraping_noKH,Ji_2019MNRAS_RMHDmixinglayers}, thereby reducing mixing.  Indeed, some observations argue for a correlation between magnetic draping and overall gas survival in stripped tails \citep{Muller_2021_RPSmagfields,Ignesti_2026arXiv260221821I_supersonicjellyfish}. However, although magnetic fields change the cloud morphology, making it more filamentary, the cloud survival criterion and mass growth rates are relatively unchanged \citep{GronkeOh2020a_coldgasentrains, Li_Hopkins_2020MNRAS_cloudsurvival_with_conductionviscosity, Sparre_2020MNRAS_cloudcrushingMHD,Jennings_2023MNRAS_CCwithMHDplusconduction}.  In particular, for the high plasma $\beta$ ($>>$1) expected in the ICM, magnetic draping has little effect on cloud survival \citep{Li_Hopkins_2020MNRAS_cloudsurvival_with_conductionviscosity}.  We note that if magnetic fields affect cloud morphology, they could therefore affect the distribution of star formation, and thus impact the stellar distribution in fireballs.

We also do not include thermal conduction, but as with magnetic fields other simulation work seems to indicate conduction will not strongly affect cloud evolution.  Simulations with isotropic Spitzer conduction do not show dramatic changes in cloud crushing simulations \citep{Bruggen_2016_outflowclouds_conduction, Armillotta2016_diskcorona}, although \citet{Armillotta_2017_cloudcrushingCGM_conduction} highlights that the distribution of cold gas is different with conduction even if the total cold mass is similar whether or not conduction is included. In addition, with more realistic anisotropic conduction there is little effect on cloud evolution because magnetic fields suppress conduction \citep{Komarov_2018JPlPh_PICconductionsupression, Li_Hopkins_2020MNRAS_cloudsurvival_with_conductionviscosity, Jennings_2023MNRAS_CCwithMHDplusconduction}.  Although simulations agree that the overall effect on cloud evolution is small, some simulations find that anisotropic conduction will enhance cloud growth \citep{Jennings_2023MNRAS_CCwithMHDplusconduction}, while others find it will destroy clouds \citep{Li_Hopkins_2020MNRAS_cloudsurvival_with_conductionviscosity}. Therefore here we do not make any predictions for the impact on fireballs.  \citet{Li_Hopkins_2020MNRAS_cloudsurvival_with_conductionviscosity} also test the impact of anisotropic viscosity on cloud survival, finding it able to only slightly increase the lifetime of some clouds via the formation of a viscous `boundary layer'. 

Including radiative transfer directly in our simulation could also have an effect on the survival of our main clump as well as the shredded material streaming along the wind direction.  Right now, the analytic prescription determining the level at which self-shielding attenuates the photoionization and photoheating rates is calculated for each cell independently.  However, we would expect that gas at the center of larger and more massive clumps would in fact be more heavily shielded by the surrounding material.  Therefore, we could be underestimating self-shielding and more gas may survive and be able to form stars within the shredded cloud material.  This would be very interesting to test using a full radiative-hydrodynamics simulation.

We finally note that our main cloud is well-resolved according to cloud-crushing simulations including radiative cooling, which generally find that cloud evolution is converged at eight cells per cloud radius \citep{GronkeOh2020a_coldgasentrains, Kanjilal_2021MNRAS_cloudgrowth_mixingcooling, Tan_2021_turbulentmixinglayers_resolution}.  Our clouds begin with more than 128 cells across their radii.  We also ran our simulation using one less level of refinement, to 3 pc, and found that the mass of cold gas (T $<$ 3 $\times$ 10$^4$ K) is very similar (within about 20\%) throughout the simulation.  Lower resolution does result in a smaller amount of gas above the star formation density threshold ($\rho$ $>$ 100 cm$^{-3}$), however, as well as a lower peak star formation rate - although the star formation rate falls off less steeply than the {\fid} run.  In total about half as many stars form in the lower resolution simulation.  However, our result of an overall age gradient in the main clump holds, as well as that most stars form within a main clump with only a small amount of star formation in the extended material that is younger than the main stellar clump.

\section{Conclusion}\label{sec:conclusion}

In this paper we have run a suite of wind-tunnel simulations of cold gas clouds interacting with a hot surrounding wind, mimicking the evolution of gas clouds in the tails of ram pressure stripped galaxies.  We have focused on the age gradient of stars formed during this interaction, to determine the physical processes driving the stellar age distribution observed in jellyfish tails, and whether the empirically-derived fireball model of decreasing age with increasing height along the wind direction ($z$ position in this paper) is universal.  Our main results are as follows:

\begin{itemize}
    \item In our simulations, the cloud in a static surrounding medium ({\NW}) forms the most stars.  Comparing only the clouds hit by a wind, we find that our sampled wind parameters have little effect on the final stellar mass, but a lower-density, larger cloud forms significantly less stellar mass (Table \ref{tab:clouds} \& Figure \ref{fig:HM_SFRs}).
    
    \item The vast majority of stars form in a ``main clump'' of gas that survives largely intact for tens of Myrs in most simulations.  For runs with higher ram pressure ({\fid} and {\fwind}) or lower cloud density ({\ldcloud}) the stellar distribution in the wind direction in the main clump has a broader distribution of older stars and younger stars tending to reside at higher $z$ positions, overall reflecting a fireball prediction, although not the monotonicity of the age gradient in simple illustrations (Fig \ref{fig:fireball_illustration}).  For lower ram pressure ({\swind} and {\ldwind}), the mean stellar age does not show a gradient with $z$ position in the main clump. (Figs \ref{fig:projections} - \ref{fig:age_z_overplot}).
    
    \item Following the toy fireball scenario, stars formed at higher $z$ positions along the wind direction tend to have higher velocities (Figures \ref{fig:zhist_fid} \& \ref{fig:zhists_comp}), as dense gas continues to be accelerated by the ICM wind.  In the main stellar clump, higher $z$ position and velocity is also roughly correlated with later formation times (younger stars).  The gradient in formation velocity results in more extended stellar clumps with greater differences in the mean $z$ positions of the old versus the young stellar populations at later times (Figs \ref{fig:age_z_overplot}).
    
    \item Because the cloud does not remain as a monolithic clump, but is shredded and extended along the wind direction, SF can occur throughout our entire 25 kpc box.  Although the majority of stars are found in the main clump, particularly at later times stars form throughout the extended cloud material.  Indeed, for tens of Myr, dense cloud material is found over a range of about 20 kpc in all of our simulations and therefore stars can form concurrently at $z$ positions that vary by several kpcs, meaning there is no monotonic age gradient with height when looking at stars formed from all the cloud material in the box (Figs \ref{fig:zhist_fid} \& \ref{fig:zhists_comp}).
    
    \item Although the gas cloud is accelerated by the surrounding gas, stars are unaffected by this hydrodynamical interaction.  However, the self-gravity of the gas cloud and/or the formed stars can homogenize the stellar velocities. We find that in simulations with lower ram pressure, {\swind} and {\ldwind}, the changes in stellar velocities tend to be larger than in those with higher ram pressure (Figures \ref{fig:vzratio_hist} \& \ref{fig:vzinit_tinit_vzdiff_scatter}). We posit that this is because the main dense cloud survives for longer in low ram pressure cases, gravitationally drawing stars together.  
\end{itemize}

Based on the results of our simulations, we have gained intuition on what is required for stars forming from ram pressure stripped clouds to have a fireball age gradient (decreasing stellar age with increasing height above the disk).  First, the cloud must be accelerating rapidly enough that later stars form displaced to higher positions and velocities in the wind direction, even when considering differential velocities within the cloud (for example from cloud collapse or turbulent velocities).  Second, the cloud must be shredded quickly enough that self-gravity cannot homogenize the different birth velocities or positions of the stars.  In addition, we note that when these criteria are fulfilled, the acceleration of the gas cloud and the resulting increasing stellar velocities with decreasing stellar age means that the distance between young and old stars expands with time, making the displacement between the populations more easily observed at later times.  

Importantly, we find that while our simulations do not seem to agree with the cartoon illustration of the fireball model, mainly because they do not produce a monotonic age gradient, they show several signatures of observed star forming clumps, such as a displacement in the wind direction of the mean position of young stars to the total population.  

While in this work most of our focus is on how SF within the main clump does or does not reproduce the fireball age gradient, we do see that within the shredded filamentary material SF shows clearer age gradients with height.  To test whether SF in the downwind material could be a significant source of the observed fireball population, we recommend that future work includes SF recipes of single-star formation at sub-pc resolution, and includes nonthermal pressure support in simulated clouds to delay the onset of SF after the wind onset.   

\begin{acknowledgments}

The authors would like to thank the referee for comments that improved the paper.  The simulations and analysis were run on the computing facilities at the Flatiron Institute, a part of the Simons Foundation.  ST and RS would like to thank Greg Bryan, Sean McGee, and Alessia Moretti for helpful discussions and comments. RS acknowledges financial support from FONDECYT Regular projects 1230441 and 1241426, and also gratefully acknowledges financial support from ANID – MILENIO – NCN2024$\_$112. BP acknowledges funding from the European Reasearch Council (ERC) under the Horizon 2020 Advanced Grant Programm GASP (grant agreement n. 833824).  BV acknowledges support from the INAF GO grant 2023 ``Identifying ram pressure induced unwinding arms in cluster spirals'' (P.I. Vulcani). EG acknowledges funding by the European Union – NextGenerationEU within PRIN 2022 project n.20229YBSAN – Globular clusters in cosmological simulations and in lensed fields: from their birth to the present epoch.
\end{acknowledgments}
\appendix

\section{Appendix}\label{sec:appA}

Here we present three snapshots of the full refined region of the {\fid{ simulation (starting at the cloud origination position).  We chose somewhat later times in which the dense filamentary material is extended through the entire box, and included 53 Myr because two new star particles have formed from dense material far downwind of the main clump.  Because the y-axis does not change in these panels, the movement of the main clump along the wind direction is quite clear, as well as the extension of the stars in the clump as can be seen in Figure \ref{fig:projections}.  

It is also interesting to compare the four stars downwind of the main clump at 48 Myr (with z positions ranging from 6 to 7.5 kpc) to the same stars 5 Myr later.  The youngest star is not the farthest downwind at 48 Myr, but because it is formed with a higher velocity, 5 Myr later it is the farthest downwind. Even if stars do not always form with a clean age gradient because of the constant cooling and of material accelerated from the main gas cloud, if young stars form with high velocities, they will end up farther downwind forming a classic `fireball' age gradient. 

\begin{figure*}
%    \centering
    \includegraphics[scale=0.95,trim={9mm 0mm 36.5mm 0mm},clip]{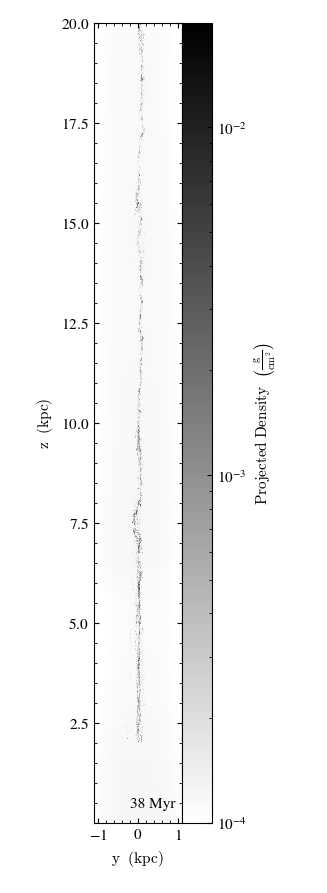}
    \includegraphics[scale=0.95,trim={23mm 0mm 36.5mm 0mm},clip]{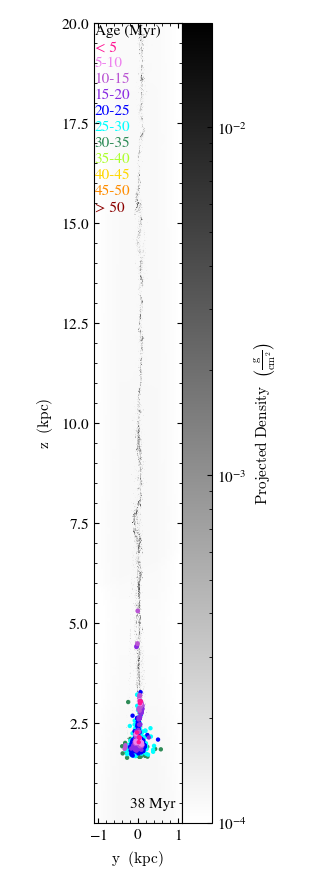}
    \includegraphics[scale=0.95,trim={23mm 0mm 36.5mm 0mm},clip]{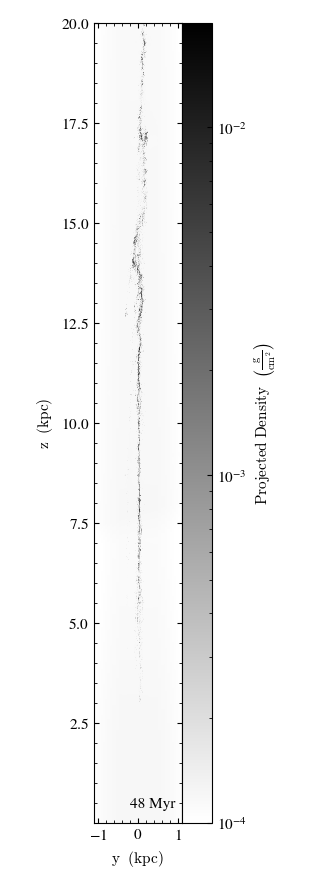}
    \includegraphics[scale=0.95,trim={23mm 0mm 36.5mm 0mm},clip]{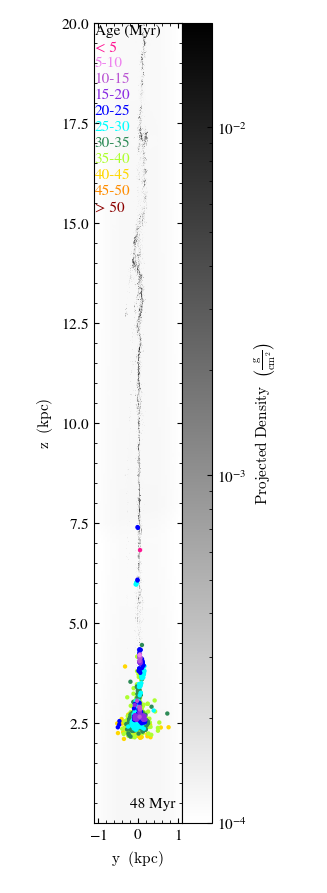}
    \includegraphics[scale=0.95,trim={23mm 0mm 36.5mm 0mm},clip]{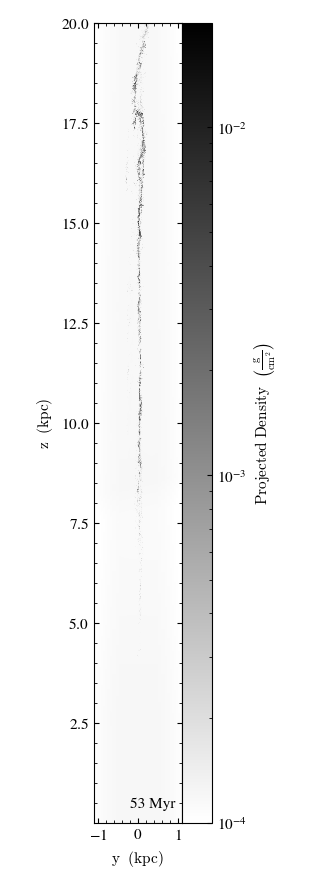}
    \includegraphics[scale=0.95,trim={23mm 0mm 13mm 0mm},clip]{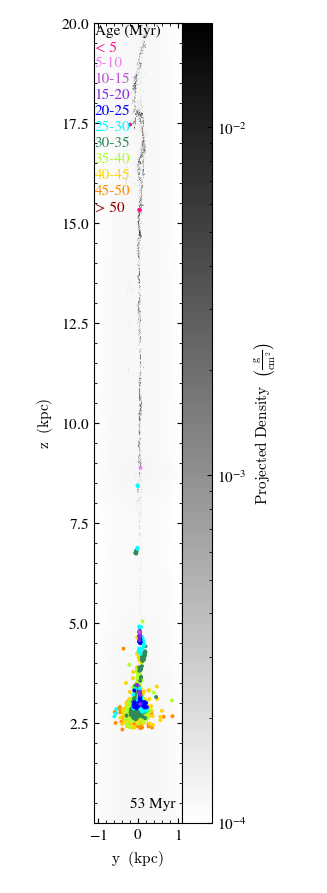}\\
    
    \caption{Three snapshots of the full refined region of the {\fid} simulation (with z=0 denoting the cloud origination position).  Each pair of outputs shows only the gas projected density (left) and the gas density with star particles overlayed (right).}
    \label{fig:projections22kpc}
    
\end{figure*}

\bibliography{references_longlist}{}
\bibliographystyle{aasjournal}

\end{document}